\documentclass[aps,twocolumn,prb,showpacs,balancelastpage,amssymb,groupedaddress]{revtex4}

\usepackage{graphicx}
\usepackage{color}

\usepackage{amssymb}
\usepackage{amsmath}
\usepackage{amscd}
\usepackage{bm}

\newcommand{\pll}{\parallel}

\newcommand{\e}{{\rm e}}
\newcommand{\rmd}{{\rm d}}
\newcommand{\rmi}{{\rm i}}

\newcommand{\tr}{\hbox{tr}}

\newcommand{\half}{{\textstyle{\frac{1}{2}}}}

\newcommand{\al}{\alpha}

\newcommand{\de}{\delta}

\newcommand{\eps}{\epsilon}

\newcommand{\ignore}[1]{\relax}
\newcommand{\tDone}{\tau_{\rm D1}}
\newcommand{\tDtwo}{\tau_{\rm D2}}

\newcommand{\tEo}{\tau_{\rm E}^{\rm op}}
\newcommand{\tEc}{\tau_{\rm E}^{\rm cl}}

\newcommand{\Nm}{N_m}
\newcommand{\Nmo}{N_{m_0}}
\newcommand{\Nmp}{N_{m'}}
\newcommand{\rhom}{\rho_m}
\newcommand{\rhomo}{\rho_{m_0}}
\newcommand{\rhomp}{\rho_{m'}}

\newcommand{\pF}{p_{\rm F}}
\newcommand{\lF}{\lambda_{\rm F}}
\newcommand{\EF}{E_{\rm F}}

\begin{document}

\title{
Suppression of weak-localization (and enhancement of noise) \\
by tunnelling in semiclassical chaotic transport}
\author{Robert S. Whitney }

\affiliation{Institut Laue-Langevin,
6, rue Jules Horowitz, BP 156,
38042 Grenoble, France.}

\date{June 5, 2007}

\begin{abstract}
We add simple tunnelling effects and ray-splitting into the recent
trajectory-based semiclassical theory of quantum chaotic transport. 
We use this to derive the weak-localization correction to conductance 
and the shot-noise for a quantum chaotic cavity (billiard) coupled to 
$n$ leads via tunnel-barriers. 
We derive results for arbitrary tunnelling rates and arbitrary (positive)
Ehrenfest time, $\tau_{\rm E}$.
For all Ehrenfest times, we show that the shot-noise is enhanced 
by the tunnelling, while the weak-localization is {\it suppressed}.
In the opaque barrier limit 
(small tunnelling rates with large lead widths, 
such that the Drude conductance remains finite), 
the weak-localization goes to {\it zero} linearly with the tunnelling rate, 
while the Fano factor of the shot-noise remains finite 
but becomes {\it independent} of the Ehrenfest time.   
The crossover from RMT behaviour ($\tau_{\rm E}=0$) to classical behaviour 
($\tau_{\rm E}=\infty$) goes exponentially with the ratio 
of the Ehrenfest time to the paired-paths survival time.
The paired-paths survival time varies
between the dwell time  (in the transparent 
barrier limit) and half the dwell time (in the opaque barrier limit).  
Finally our method enables us to see the physical origin
of the suppression of weak-localization; it 
is due to the fact that tunnel-barriers  ``smear'' 
the coherent-backscattering peak over
reflection and transmission modes.
\end{abstract}


\maketitle


\section{Introduction.}
\label{sect:intro}

\begin{figure}
\begin{center}
\hspace{-1cm}
\resizebox{4.5cm}{!}{\includegraphics{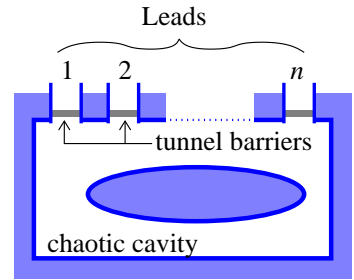}}
\caption{\label{fig:cavity} 
A chaotic cavity (billiard) with $n$ leads.
The $m$th lead has width $W_m$, and is coupled to the cavity through a 
tunnel-barrier with transmission probability, $\rho_m$.  
We assume that all the tunnel-barriers
are very high and narrow, so that this transmission probability is
independent of the lead mode number. 
}
\end{center}
\end{figure}

Semiclassical trajectory-based methods have long been applied 
to quantum chaos \cite{Haake-book,Ric-book}, however
recent years have seen phenomenal progress in this field.
The challenge of going beyond the Berry
diagonal approximation~\cite{Berry-diag} has finally been 
overcome\cite{Sieber01}.
Some works\cite{Sieber01,Ric02,Haake-levelstat,Haake-weakloc,Haake-fano,  
bsw,Schanz-fano}
have made progress towards a microscopic foundation
for the well-established conjecture
\cite{BGS} that hyperbolic chaotic systems have 
properties described by random matrix theory (RMT).
Other works\cite{Ale96,agam,Ada03, wj2004,Brouwer-quasi-wl,wj2005-fano,jw2005, 
Brouwer-cbs,Brouwer-ucf,Brouwer-quasi-ucf,Brouwer-noise}
have explored the crossover from the RMT regime 
(negligible Ehrenfest time) to the 
classical limit (infinite Ehrenfest time).
Most of the latter (and some of the former) 
works have dealt with mesoscopic effects in the 
quantum transport of electrons through open quantum dots with
chaotic shape.  They have analyzed the weak-localization 
correction to conductance and the resulting magneto-conductance
\cite{Ale96,Ric02,Ada03,Haake-weakloc,Brouwer-quasi-wl,jw2005,Brouwer-cbs},
the shot-noise (intrinsic quantum noise)
\cite{agam,Haake-fano,wj2005-fano,Brouwer-noise},
and conductance fluctuations
\cite{Brouwer-ucf,Brouwer-quasi-ucf}.
However to-date all such works on quantum transport have 
assumed perfect coupling to the leads.
The objective of this work is to consider
quantum transport through a chaotic system that is coupled to
the leads via tunnel-barriers (see Fig.~\ref{fig:cavity}).
Such systems are experimentally realizable as large ballistic 
quantum dots\cite{marcus}. 
We study two properties of quantum transport, which highlight 
different aspects of the wave-nature of electrons.
The first is the weak-localization correction to conductance,
it is a contribution to conductance due to interference between 
electron-waves.
It is destroyed by a weak-magnetic field, leading to 
a finite magneto-conductance. The second is the shot-noise,
an intrinsically quantum part of the fluctuations of 
a non-equilibrium electronic current,
it is due to the fact that different parts of the electron-wave can 
go to different places.

If one assumes the
chaotic system is well-described by a random matrix from the appropriate 
random matrix 
ensemble, then it is known how to calculate transport through a 
chaotic system coupled to leads via 
tunnel-barriers\cite{Iida90-RMT-transport,Brouwer96-RMT-transport-diagrams}.
Under this assumption it has been shown that a cavity 
with two identical leads (both with $N$ modes and with barriers
with tunnelling probability $\rho$),
has a Drude dimensionless conductance 
$g^{\rm Drude} = \rho N/2$, while the weak-localization
correction goes like $g^{\rm wl} = -\rho/4$.
Thus the weak localization correction goes to zero when we take $\rho \to 0$, 
even when one keeps the Drude conductance constant (by taking $N\to \infty$
such that $\rho N$ remains constant).
The formal nature of the RMT derivations have made it hard to
give a simple physical explanation of why this suppression of weak-localization
occurs.  Indeed the suppression seems counter-intuitive, since 
we know that in a chaotic (ergodic and mixing)
system the details of the system dynamics 
have little effect on the transport properties.
Thus one would expect that transport through a lead with 
such tunnel-barriers should be equivalent to transport through a 
lead with $N'=\rho N$ modes without tunnel-barriers.  
This argument predicts the correct Drude conductance, 
however it would lead one to predict {\it incorrectly} that the
weak-localization correction is $-1/4$, instead of $-\rho/4$.
The vanishing of weak-localization in the limit of strong tunnelling
is intriguing because it shows that there can be competition
between two of the most fundamental aspects of quantum mechanics:
tunnelling and interference.

One problem with the random matrix assumption is
that it tells us nothing about the cross-over to the classical limit.
The parameter which controls this crossover is the ratio of the Ehrenfest
time to a dwell time.
The Ehrenfest time (first introduced into chaotic quantum transport in 
Ref.~[\onlinecite{Ale96}]) is given by the time it takes for 
a wavepacket which is well-localized in phase-space
to gets stretched to a classical lengthscale.
Since the wavelength, $\lF=\hbar/\pF$, 
is much smaller than the system size, $L$, the stretching
is well approximated by the flow of the classical dynamics.
In a hyperbolic chaotic system this stretching is exponential, and happens
at a rate given by the Lyapunov exponent, $\lambda$.
It is convenient to define the Ehrenfest time as twice the time
for a minimal wavepacket (spread by $(\lambda_{\rm F}L)^{1/2}$ 
in position and $(\lambda_{\rm F}L/\hbar)^{-1/2}$ in momentum)  
to stretch to a classical scale.
Typically in open chaotic systems there are two classical scales,
the system size $L$, and the lead width $W$;
there is an Ehrenfest time associated with each 
scale\cite{Vavilov,Scho05},
\begin{eqnarray}
\tEc &=&\lambda^{-1} \ln [L/\lF], \\
\tEo &=& \lambda^{-1} \ln [(L/\lF)(W/L)^2].
\end{eqnarray}
The first we call the {\it closed-cavity} Ehrenfest time
since it is the only Ehrenfest scale in a closed system,
the second we call the {\it open-cavity} Ehrenfest time, because it is
associated with the opening of the leads.  However we emphasis
that both lengthscales ($L$ and $W$), and hence both Ehrenfest times, 
are relevant to open cavities\cite{wj2005-fano,jw2005}.

On times shorter than the Ehrenfest times the dynamics of the system are
not sufficiently mixing to give RMT results for transport
quantities.  In this article we address for the first time
how tunnelling affects transport when a significant proportion of the
current is carried by paths shorter than the Ehrenfest times.
To do this we must include the ray-splitting caused by such tunnel-barriers
into the semiclassical trajectory-based method.
We then
investigate the effect of tunnelling on weak-localization
and intrinsic quantum noise (shot noise)
in a system with finite Ehrenfest time.
In addition to this we find that the method enables us
to clearly see the physical mechanism
by which tunnelling suppresses the weak-localization effects.  
The method presented here was used in Ref.~[\onlinecite{petitjean06}]
to investigate the effect of dephasing voltage probes on 
weak localization.

This article is organized as follows.
Section \ref{sect:qualitative}
contains qualitative explanations of the central results in this article
(particularly the 
suppression of weak-localization and Ehrenfest-independence of the shot-noise
in the opaque barrier limit). 
Section \ref{sect:classical} 
discusses the dynamics of the classical paths which are relevant for
the semiclassical calculations.
Section \ref{sect:semiclass}
sets up the formalism for including tunnelling in the semiclassical 
trajectory-based method.   Sections \ref{sect:weak-loc} 
and \ref{sect:noise} then present calculations of the 
weak-localization correction 
to conductance and the Fano factor of the shot-noise, respectively.

\section{Qualitative discussion of results.}
\label{sect:qualitative}

\subsection{Suppression of
weak-localization by the smearing of the coherent-backscattering peak.}
\label{sect:failed-cbs}

As noted in \cite{jw2005,Brouwer-cbs} for a system without tunnel-barriers, 
the coherent-backscattering
approximately doubles the weight of all paths returning 
close and anti-parallel to the same path at injection.
The approximate doubling of the weight applies to all
paths that return to a strip defined by $(y,\theta)$ across the lead, where
\begin{eqnarray}
\theta-\theta_0 \simeq - p_{\rm F}^{-1} m \lambda (y-y_0) \cos \theta_0 \ 
\end{eqnarray}
for a path that was initially injected at $(y_0,\theta_0)$. 
This strip sits on the stable axis of the classical dynamics,
with a width in the unstable direction of order 
$\hbar(p_{\rm F}W)^{-1}$.
The probability for the particle to go anywhere else is slightly reduced
(by the weak-localization correction which is negative and 
$(y,\theta)$-independent), 
this ensures that the
probability for the particle to go somewhere remains one 
(preserving unitarity and conserving current).
Since the enhancement due to coherent-backscattering only occurs for
reflection, we see that reflection is slightly 
enhanced and transmission (and hence
conductance) is slightly reduced; this is the weak-localization correction
to conductance.

However once we introduce tunnel-barriers the situation becomes more
complicated.  There is still the enhancement of paths that return to 
close and anti-parallel to themselves at injection, however there is
no longer a guarantee that this enhancement contributes to reflection.
The enhanced paths can return to the injection lead but then be reflected
off the tunnel-barrier, remaining in the cavity (see Fig.~\ref{fig:wl}c).
These enhanced paths will then bounce in the cavity until they
eventually transmit through a barrier, either contributing to transmission
(as shown in Fig.~\ref{fig:wl}c) or reflection
 (as shown in Fig.~\ref{fig:wl}c if we set $m=m_0$).
We use the term 
{\it successful coherent-backscattering} 
to refer to the the usual coherent-backscattering contribution,
shown in Fig.~\ref{fig:wl}b, while using the term
{\it failed coherent-backscattering} for those contributions
which behave in the manner shown in Fig.~\ref{fig:wl}c,d. 

Thus the failed coherent-backscattering 
gives a positive contribution to both reflection and transmission
which will tend to off-set the negative one coming from the usual
weak-localization correction.  This contribution will go like 
$(1-\rho)$ times the coherent-backscattering contribution in the absence
of tunnel-barriers.
This contribution will be ``smeared'' over all leads
with a weight for each lead equal to the probability of escaping 
through that lead.  So if the cavity has two identical leads, 
half of it will contribute to transmission and the other half to reflection. 
The successful coherent-backscattering
goes like $\rho$ and only contributes to reflection.

As we take $\rho \to 0$ 
we see that the coherent back-scattering becomes 
completely ``failed'' (the successful part goes to zero).
So the whole backscattering peak is smeared over all leads
in exactly the same way as the conventional weak-localization correction.
Both have a weight for any given lead equal to the probability for
the particle to escape through that lead.
However the two have opposite signs (which they must to preserve unitarity) 
and thereby cancel.  Hence the total weak-localization correction  
(sum of the conventional weak-localization correction, Fig.~\ref{fig:wl}a,
and the failed coherent-backscattering, Fig.~\ref{fig:wl}c,d) 
vanishes when we take $\rho \to 0$
(even if we keep $\rho N$ finite, so the Drude
conductance remains finite).

\subsection{RMT to classical crossover given by ratio
of Ehrenfest time to paired-paths survival time.}

In all calculations in this article we find that
the crossover from RMT behaviour (zero Ehrenfest time)
to classical behaviour (infinite Ehrenfest time)
is governed by the ratio of the Ehrenfest time
to the {\it paired-paths} survival time, not the {\it single-path}
survival time (the dwell time).
The paired-paths survival time, defined in Eq.~(\ref{eq:P2}), is the
time during which a pair of initially identical classical paths both remain
inside the cavity.  Under the classical dynamics
of a closed cavity a pair of initially identical paths remain paired 
(identical) forever.  
However the presence of the leads means that all paths
remain a finite time in the cavity, we call this the dwell time
or single-path survival time, $\tDone$. 
Further, if there are tunnel-barriers
on the leads then the paired-paths survival time, $\tDtwo$, 
is given by the time for one or both of two (initially identical) 
paths to escape.  We show that this time 
varies between $\tDone$ for transparent barriers to $\half\tDone$ for 
opaque barriers.

The reason why $\tDtwo$ is the relevant timescale for the suppression
of the crossover is as follows. All paths which contribute to
the RMT limit must involve an intersection at an angle of order
$(\lambda_{\rm F}/L)^{1/2}$, and the paths must the survive in the cavity until
they spread to a distance apart of order the width of the lead
on both sides of the intersection.
Within $\half \tEo$ on either side of the intersection the paths
are correlated, and so their joint survival probability decays at the rate 
$\tDtwo^{-1}$. 
The shot noise contributions divide into those in which 
the paired-paths survive for a time less than $\tEo$   
and those which survive a time greater than $\tEo$.
Thus we see that these contributions go like $(1-\exp[-\tEo/\tDtwo])$ and
$\exp[-\tEo/\tDtwo]$ respectively.

All weak-localization contributions require
that the path segments diverge to a distance of order the system size
on one side of the intersection (so that a closed loop can form) and to
a distance of order the lead width on the other side 
(when the ``legs'' can escape).
Thus the path segments must survive until a time $\tEc$ on the loop-side of
the intersection and $\tEo$ on the leg-side.  
However within $\tEo/2$ on both sides of the intersection,
the paths have the paired-paths survival probability 
(because they are closer than the lead width), while elsewhere
each path segment individually has the single-path escape
probability.  Thus we see that the exponential suppression 
of weak-localization for finite Ehrenfest time goes like
the survival probability for such paths of length $\tEo+\tEc$
which is
\begin{eqnarray}
\exp [-\tEo/\tDtwo - (\tEc-\tEo)/\tDone].
\end{eqnarray}
The first term in the exponent is the Ehrenfest time dependence,
the second term is a classical correction, since
$(\tEc-\tEo)$ is independent of the particle wavelength.
In the transparent barrier limit  (where $\tDtwo=\tDone$)
the suppression goes like $\exp [-\tEc/\tDone]$ as found in
Ref.~[\onlinecite{jw2005}], while
in the opaque barrier limit we find $\tDtwo=\tDone/2$
so the suppression goes like $\exp [-(\tEc+\tEo)/\tDone]$.
  
In general the survival probability of such intersecting paths 
of total length $t>2\tEo$ is 
\begin{eqnarray}
\exp[-(t-2\tEo)/\tDone - \tEo/\tDtwo],
\end{eqnarray}
since the paths are paired-during a time $\tEo$ and unpaired at all other times
(when unpaired each segment decays at a rate of $\tDone^{-1}$). 
In the transparent barrier limit
this is $\exp[-(t-\tEo)/\tDone]$
while in the opaque barrier limit it is simply
$\exp[-t/\tDone]$.

\subsection{Ehrenfest time independence of shot noise in the opaque
barrier limit.}

In the limit of opaque barriers
(small tunnelling rate)
we find that the shot noise is independent of the
Ehrenfest time up to first order in $\rho$, see Eq.~(\ref{eq:Fano-opaque2}).
Thus for any small $\rho$, the magnitude of the difference between the noise
in the RMT limit ($\tEo = 0$) and the classical limit ($\tEo=\infty$)
goes like $\rho^2$, 
and thus may be beyond experimental resolution.
This is despite the fact that the contributions 
in the RMT limit are very different from those in the classical limit.  
Closer inspection shows that those
contributions which survive in the opaque limit
(Fig.~\ref{fig:noise-trans1}e-i and 
Fig.~\ref{fig:noise-trans2}b-g)
do have one thing in common;
none of them have correlations between the paths when entering or 
exiting the cavity.
Thus it appears that it is irrelevant 
whether correlated paths have enough
time to become
uncorrelated inside the cavity (as they do when $\tEo \ll \tDtwo$, but
not when $\tEo \gg \tDtwo$) because the tunnel-barriers destroy all 
correlations between paths (the probability that both paths exit 
when they hit the tunnel-barrier is zero in this limit).
It is intriguing that the shot noise is so insensitive
to the manner in which correlations between paths are destroyed
(either via classical dynamics or via tunnelling).

We note that for a symmetric two-lead system 
in the opaque limit ($N_1=N_2$ and $\rho_1=\rho_2\ll 1$),
the shot noise is half that for Poissionian noise
(Fano factor, $F=1/2$).  
Thus even through the transmission probability is small, 
so transmission is rare, the transmission is sub-Poissonian
process.  
The Fermionic nature of the electrons induces correlations
even in this limit.
The result that $F=1/2$, is the same as one would
get for two leads coupled by a single tunnel-barrier with transmission
probability equal to $1/2$ (with no chaotic system present at all).


\section{Classical dynamics in a system with tunnel-barriers on the leads.}
\label{sect:classical}

In this article we consider a quantum system whose classical limit
(wavelength $\lambda_{\rm F} \to 0$) has a non-zero tunnelling
probability at tunnel-barriers (we assume the width of the barriers
scales with $\lambda_{\rm F}$).
Thus it is natural to include tunnelling at the level of the classical
dynamics.  
Here we introduce tunnelling into the classical dynamics phenomenologically,
we will then see in section~\ref{sect:semiclass} 
that it is the classical limit of the quantum problem we wish to solve. 
We assume that each time a classical path hits a barrier
there is a probability of $\rho$ that the path goes through the barrier
(as if the barrier is absent) and a probability of $(1-\rho)$
that the particle is specularly reflected (as if the barrier is impenetrable).

The probablistic nature of this tunnelling 
makes the classical dynamics {\it stochastic}.  The classical
paths, which are the solutions of these stochastic dynamics,
have properties not present in deterministic dynamics,
such as bifurcations (ray-splitting) 
at the tunnel barriers 
(one part transmitting and the other reflecting).
To simplify the considerations in this article
we consider tunnel-barriers that have a  
tunnelling probability which is independent of the angle, $\theta$, 
at which a classical path hits the barrier.
This assumption is justified for tunnel-barriers in the limit 
of large barrier height and small barrier thickness 
(see Appendix ~\ref{sect:barrier}).
This is equivalent to saying
we consider that the tunnel-barriers have the same tunnel probability
for all lead modes.  

Let us now consider the dynamics of a classical particle in such a system.
We define ${\tilde P}({\bf Y},{\bf Y}_0;t)$ such that
the classical probability for a particle to go from 
an initial position and momentum angle of
${\bf Y}_0\equiv(y_0,\theta_0)$ 
in the phase-space of lead $m_0$ (a point on the cross-section of the lead
just to the lead side of the tunnel-barrier) to within
$(\de y,\de \theta)$ of 
${\bf Y}=(y,\theta)$ 
on the cross-section 
of lead $m$ (again just to the lead side of the tunnel-barrier) 
in a time within $\de t$ of $t$ is 
\begin{eqnarray}
{\tilde P}({\bf Y},{\bf Y}_0;t)\de y\de \theta \de t .
\end{eqnarray}
In a system {\it without} tunnel-barriers the classical dynamics are 
deterministic, and ${\tilde P}({\bf Y},{\bf Y}_0;t)$ 
has a Dirac $\de$-function with unit weight on classical paths.
However in a system {\it with} tunnel-barriers the classical dynamics 
are stochastic, with each barrier acting as a {\it ray-splitter}. 
A classical path which hits a barrier is split into two
(one which passes through the barrier and one which reflects).
Thus,
${\tilde P}({\bf Y},{\bf Y}_0;t)$
has a $\de$-function on each classical path which exists (counting all
possible ray-splittings), however the weight of the Dirac $\de$-function is
the product of the tunnelling/reflection probabilities
that that path has acquired each time it has hit a tunnel-barrier.
Thus the integral of ${\tilde P}({\bf Y},{\bf Y}_0;t)$ 
over all positions/momenta at escape, ${\bf Y}$,
gives a sum over all paths starting from ${\bf Y}_0$ in which each term
is weighted by these tunnelling/reflection probabilities, hence
\begin{eqnarray}
& &\hskip -5mm
\int_0^\infty \rmd t\int \rmd {\bf Y} {\tilde P}({\bf Y},{\bf Y}_0;t) 
\; [\cdots]
_{{\bf Y}_0}
\nonumber \\
&=& \sum_{\gamma \in \{{\bf Y}_0\} }
\rho_m \rho_{m_0}
{\textstyle \prod_{m'}}  \big[1-\rho_{m'}\big]^{n_\gamma (m')}  
\; [\cdots]_\gamma
\label{eq:integral-as-path-sum}
\end{eqnarray}
where $\rmd {\bf Y}= \rmd y \rmd \theta$, with $y$ integrated 
over the cross-section on each lead and $\theta$ integrated from 
$-\pi/2$ to $\pi/2$.
The sum is over all possible path starting at 
${\bf Y}_0 \equiv (y_0,p_{y_0})$. 
Lead $m$ is defined as the lead the path finally escapes into, and
$n_\gamma (m')$ is the number of times that the path $\gamma$ reflects
off the tunnel-barrier on lead $m'$ before escaping. 
We assume that any quantities in $[\cdots]_\gamma$ are  
independent of $\rho_{m'}$, in-other-words they are 
same as for the path which would exist if the tunnel-barriers
were impenetrable for each reflection and absent for each tunnelling
of path $\gamma$.

In semiclassics one is typically summing over all paths, $\gamma$,
from $y_0$ to $y$ with energy, $E$, rather than 
all paths starting at ${\bf Y}_0$.
Using Eq.~(\ref{eq:integral-as-path-sum}) 
we see that this sum can be written in terms of 
 ${\tilde P}({\bf Y},{\bf Y}_0;t)$ as
\begin{eqnarray}
& & \hskip -10mm
\sum_{\gamma \in \{y_0\to y;E\} } \rho_m \rho_{m_0}
{\textstyle \prod_{m'}}  \big[1-\rho_{m'}\big]^{n_\gamma (m')} 
\; [\cdots]_\gamma
\nonumber \\
&=& \int_0^\infty \rmd t \int_{-\pi/2}^{\pi/2} \rmd \theta_0 \rmd \theta 
\left(\rmd y \over \rmd \theta_0 \right)  \tilde{P}({\bf Y},{\bf Y}_0;t)
\; [\cdots]_{{\bf Y}_0} \quad 
\label{eq:path-sum-as-integral}
\end{eqnarray}
where the factor of $(\rmd y/\rmd \theta_0)$ comes from the change of 
integration variable from $y$ to $\theta_0$.

\subsection{Ensemble or energy average of classical probabilities.}

If we average, $\langle \cdots \rangle$, 
over energy or an ensemble of similar chaotic systems, 
the average of ${\tilde P}({\bf Y},{\bf Y}_0;t)$ will be a smooth function.
If the system is mixing (when the leads are absent) 
and we perform sufficient averaging, we can assume the probability 
to go to any place in the cavity phase space is uniform.
From this it immediately follows that 
the probability to go from a point in lead $m_0$ to anywhere in lead $m$
(for $m\neq m_0$) is
\begin{eqnarray}
\int_0^\infty \rmd t\int_m \rmd {\bf Y} \langle 
{\tilde P}({\bf Y},{\bf Y}_0;t) \rangle
&=&
\rhomo \times {\rhom W_m \over \sum_{m'} \rhomp W_{m'}}
\nonumber \\
&=& {\rhomo \rhom \Nm \over \sum_{m'} \rhomp \Nmp},
\label{eq:average-trans-prob}
\end{eqnarray}
where ${\bf Y}$ is integrated over the cross-section of lead $m$. 
To see this we simply note that the probability to enter the cavity from 
lead $m_0$ is $\rho_{m_0}$, and the probability to escape the cavity is one.
Thus the probability to escape 
into lead $m$ is the ratio of $\rhom W_m$ to the sum of $\rhomp W_{m'}$
over all leads.
Note that in doing this we ignore all unsuccessful attempts to escape,
thus the particle may have hit tunnel-barriers on the leads
many times, but each time been reflected.
The result in Eq.~(\ref{eq:average-trans-prob})
can also be derived by noting that the probability 
of hitting the tunnel-barrier on lead $m_i$ is $W_{m_i}/(\tau_0 L)$ 
where $\tau_0$ is the time of flight
across the cavity, and then explicitly 
summing all reflections off the barriers.
However we feel the above logic of ignoring all unsuccessful escape attempts
gives the results in a direct and simpler manner,
thus we use it throughout this article.

Following the same logic that leads to
Eq.~(\ref{eq:average-trans-prob}), 
we see that the survival probability of a single classical path which
is inside the cavity 
is given by the master equation,
\begin{eqnarray}
\dot P_1(t) = - \tDone^{-1} P_1(t)
\label{eq:P1}
\end{eqnarray}
where we define the single path dwell time as
\begin{eqnarray}
\tDone^{-1} = (\tau_0 L)^{-1}\sum_{m=1}^n \rho_mW_m.
\label{eq:tDone}
\end{eqnarray}
Thus the probability to tunnel into the cavity from lead $m_0$ then escape
into lead $m$ at position ${\bf Y}=(y,\theta)$ on the phase-space cross-section
just to the lead side of the tunnel-barrier is
\begin{eqnarray}
\langle {\tilde P}({\bf Y},{\bf Y}_0;t) \rangle 
= \frac{\rho_{m_0}\rho_m \cos \theta }
{2 (\sum_{m'} \rho_{m'} W_{m'}) \tDone} \; 
\exp[-t/\tDone] \, .
\label{eq:average-tildeP}
\end{eqnarray}

Now we turn to the evolution of a pair of almost identical paths.
A quick glance at the figures in this article show that such pairs of
paths (vertically cross-hatched regions in 
Figs.~\ref{fig:encounter}-\ref{fig:noise-refl2})
are crucial to the quantities we calculate. 
This situation is different from the evolution of two very different paths,
because here the two
paths explore the same regions of the cavity's phase-space. 
Thus if one path hits a tunnel-barrier then so will the other one.
This leads to the definition of ``almost identical'';
the paths must be a perpendicular distance apart that is less than
the lead width for a significant period of time (at least a few bounces),
i.e. their difference in momenta
is less than $\pF \times W/L$. 
(In principle two paths with very different momenta, will 
be ``almost identical'' very close to the points where they cross,
however we ignore this as it will not significantly affect our calculations.)  
Then when the pair of paths hit a tunnel-barrier the 
probability to survive (probability for both paths to remain in the cavity) 
is $(1-\rho_m)^2$, this means the probability to 
{\it not} survive is $\rho_m(2-\rho_m)$.
Thus the survival probability of paired-paths is
given by the master equation 
\begin{eqnarray}
\dot P_2(t) = - \tDtwo^{-1} P_2(t)
\label{eq:P2}
\end{eqnarray}
where  we define the paired-paths survival time as
\begin{eqnarray}
\tDtwo^{-1} = (\tau_0 L)^{-1}\sum_{m=1}^n \rho_m(2-\rho_m)W_m.
\label{eq:tDtwo}
\end{eqnarray}
Note that this is the probability that both paths survive.  
Since $0\leq \rho_m \leq 1$ we see that
$1\leq \tDone/\tDtwo \leq 2$, 
with $\tDtwo=\tDone$  when all tunnel-barriers are transparent 
($\rho_m=1$ for all $m$), and $\tDtwo=\tDone/2$ 
in the limit of opaque
barriers ($\rho_m\to 0$ for all $m$). 
Unfortunately we cannot write down an equation for the 
evolution of pairs of paths which is as simple as 
Eq.~(\ref{eq:average-tildeP}), because even under averaging the
position/momentum of the second path in the pair are 
{\it deterministically} given by its initial position/momentum
relative to the first path.
Assuming the system has locally uniform hyperbolic dynamics 
(on lengthscales up to $W \ll L$), then the dynamics of the second
path in coordinates perpendicular to the first are 
\begin{eqnarray}
{\rmd \over \rmd t}
\left(\begin{array}{c} 
r_\perp/L \\ p_\perp/\pF \end{array} \right) = {\cal M}
\left(\begin{array}{c} 
r_\perp/L \\ p_\perp/\pF \end{array} \right)   
\end{eqnarray}
where ${\cal M}$ is a two-by-two matrix.
For Hamiltonian dynamics  ${\cal M}$ has eigenvalues $\pm \lambda$
where $\lambda$ is the Lyapunov exponent,
and so the dynamics are area preserving.
We will assume for simplicity that the eigenvector of ${\cal M}$ 
with eigenvalue $\pm\lambda$ is along the axis defined 
by $p_\perp =\pm  m\lambda r_\perp$.
We will use this to calculate the relative position of the second path in a 
pair when it is needed in the calculations below.


\section{Trajectory-based semiclassics with tunnel-barriers.}
\label{sect:semiclass}

The semiclassical derivation of the Drude conductance has become standard
\cite{Bar93,Ric-book} 
(here we will broadly follow Ref.~[\onlinecite{jw2005}]).
However we wish to introduce tunnelling into this derivation,
this presents a difficulty since tunnelling cannot be described in
conventional semiclassics. 
To deal with this we
follow a well-established procedure
for dealing with a semiclassical system 
(wavelength much less than other lengthscales) in which there are isolated 
regions where semiclassics fails, 
see for example Ref.~[\onlinecite{matching-semicl}].
We treat the regions where semiclassics fails in an exact
manner (or using an appropriate approximation scheme) and then couple the 
propagators in those regions with the semiclassical ones in  
the regions where semiclassics works well.

\subsection{The energy Green's function and scattering matrix elements.}
\label{sect:Energy Greens}

Before addressing the construction of an energy Green's function
which includes the tunnelling, let us consider the case of a {\it ray}
inside the cavity hitting the tunnel-barrier on one of the leads.
A ray being a plane-wave multiplied by an envelope-function in the direction 
perpendicular to its motion which is much wider than $\lambda_{\rm F}$, 
but much narrower than the classical scales, $W,L$. 
In this case we can treat the scattering of the ray 
in the same manner as a plane-wave hitting the tunnel-barrier 
on a lead of infinite width.  
The equations for
motion parallel and perpendicular to the barrier can be solved separately. 
The evolution perpendicular
to the barrier is given by the solution of a textbook one-dimensional 
tunnelling problem
(see Appendix~\ref{sect:barrier}), while the evolution 
parallel to the barrier is unchanged by the presence of the barrier.
The solution of the one-dimensional tunnelling problem tells
us that for any given ray arriving at the tunnel-barrier, 
there will be two rays leaving the tunnel-barrier.
One ray is that transmitted through the barrier, it has complex amplitude
$t$ and has the same momentum as the incoming ray. 
The other ray is that reflected off the barrier, it has complex amplitude
$r$ and has its momentum perpendicular to the barrier reversed
(specular reflection).
Thus each ray will be split into two each time it encounters
a tunnel-barrier.  Of course the amplitude of the two new rays are such that
$|r|^2+|t|^2 =1$.
To fit with the classical model in Section~\ref{sect:classical},
we define $\rho=|t|^2$.

For a very narrow barrier (which is the only case we consider in this article)
these two rays are identical to the rays
one would have if the barrier was both absent (transmitted ray) and 
impenetrable (reflected ray), 
except that here the amplitude of the rays are multiplied by 
the complex amplitudes $t$ and $r$ respectively.
For such narrow tunnel-barriers, the amplitudes $t$ and $r$ are independent
of the angle, $\theta$, of the incoming ray.

\begin{figure}
\begin{center}
\resizebox{6.5cm}{!}{\includegraphics{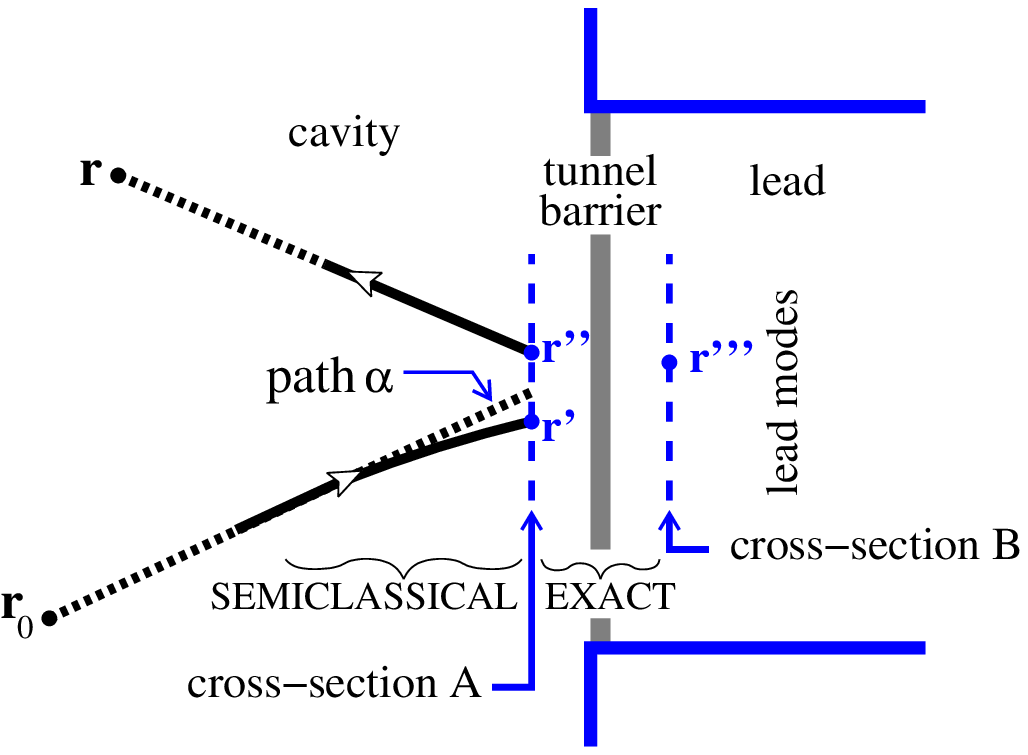}}
\caption{\label{fig:tunnel} 
Schematic of the calculation of Green's functions in the vacinity
of a tunnel barrier.  The evolution from ${\bf r}'\to{\bf r}''$
and  ${\bf r}'\to{\bf r}'''$ is treated exactly, 
while semiclassics is used for the evolution from ${\bf r}_0\to{\bf r}'$ 
and ${\bf r}''\to {\bf r}$.  
We calculate the contribution for the part of the incoming Green's function 
due to paths in the vacinity of the classical path $\al$,
by matching the semiclasical and exact solutions of the Green's function 
equation on the cross-section A, 
which is just to the left of the tunnel-barrier.
Just to the right of the tunnel-barrier, on cross-section B,
we couple to the lead modes. 
}
\end{center}
\end{figure}
 
Now we must do the same for the energy Green's function for propagating
from $r_0$ in the L lead to $r$ in the R lead.
We do this in a manner similar to Ref.~[\onlinecite{matching-semicl}].
We first note that at large distances, 
$|{\bf r}-{\bf r}_0| \gg \lambda_{\rm F}$, 
a semiclassical Green's function, $G({\bf r},{\bf r}_0;E)$, 
is well approximated by a plane-wave in the vacinity 
of any given classical path. 
Further for ${\bf r}\neq{\bf r}_0$ the differential equations for 
the Green's function and a wavefunction are the same (the Schr\"odinger 
equation).
Thus we treat the semiclassical Green's function in the vacinity of 
a given path
which touches the tunnel barrier (path $\al$), as a plane-wave or ray.
We use this as the ingoing boundary condition on cross-section A
(see Fig.~\ref{fig:tunnel}) 
for the tunnelling/reflection
problem.  The tunnelling-reflection problem is solved using the standard method
for wavefunctions (see Appendix~\ref{sect:barrier}).  The transmitted part
gains a complex prefactor, $t$, 
and then couples to the lead modes (as in the case without tunnel barriers,
we assume the leads are wide enough that they accept all momentum states).
The reflected part gains a complex amplitude, $r$, 
and is a plane-wave with its momentum perpendicular to the barrier reversed
compared with the incoming plane-wave.
Treating this outgoing plane wave, at cross-section A,
as a boundary-condition on the 
semiclassical evolution, we see it couples to exactly the same paths as if the
barrier was impenetrable.  
Hence every tunnel-barrier couples each incoming path to two outgoing paths
(one as if the barrier was absent and one as if the barrier were impenetrable).
If the weight on the incoming path is $B_\al$ then the weight 
on the two outgoing paths will be $tB_\al$ and $rB_\al$ for transmisson
and reflection at the barrier, respectively.

Hence when we write the semiclassical Green's function inside the cavity
in the usual way, it is a sum
over classical paths inside the cavity which can either transmit or reflect
at each barrier. 
The properties of the classical path 
(action, $S_\gamma$, Maslov index, $\mu_\gamma$, and classical stability) 
are the same as for the classical path
that would exist if the barrier were absent for 
each transmission and impenetrable for each reflection.
The full energy Green's function, including the tunnel-barriers,
then takes the form
$\sum_\gamma B_\gamma \exp[\rmi S_\gamma/\hbar  + \rmi \mu_\gamma/2]$, 
where $B_\gamma$ is the squareroot of the stability of the path
multiplied by a complex factor of 
$t_{m'}$ and $r_{m'}$ for each transmission and reflection at barrier $m'$. 
Using Ref.~[\onlinecite{Bar93}] to go from this Green's function 
to the formula for ${\mathbb S}_{mm_0;ji}$,
the scattering matrix element to go from mode $i$ on lead $m_0$ to mode
$j$ on lead $m$, we get
\begin{eqnarray}\label{semicl-tr}
{\mathbb S}_{mm_0;ji} 
&=&
-(2\pi \rmi \hbar)^{-1/2}
\!\int_{\rm L} \! \! \rmd y_0 \int_{\rm R} \! \rmd y 
\sum_\gamma A_\gamma
\nonumber \\
& &\qquad \times \langle j|y\rangle  
\langle y_0| i \rangle 
\exp[\rmi S_\gamma /\hbar + \rmi \pi \mu_\gamma/2 ]
\, ,\qquad
\end{eqnarray} 
where $|i\rangle$ is the transverse wavefunction
of the $i$th lead mode on lead $m'$. 
The sum is over all paths $\gamma$ (with classical action $S_{\gamma}$ 
and Maslov index $\mu_\gamma$) which start at $y_0$ on the cross-section of
the injection ($m_0$) lead, tunnel into the cavity,  bounce many times
(including reflecting off the tunnel-barriers on the leads) before
tunnelling into lead $m$ at point $y$ on its cross-section.
The complex amplitude $A_\gamma$ is  
\begin{eqnarray}
\label{eq:A}
A_\gamma &=&  
\left(\rmd p_{y_0}\over \rmd y\right)^{1/2}_{\gamma}    
t_m t_{m_0}
\prod_{m'}  \big[r_{m'}\big]^{n_\gamma (m')}
\end{eqnarray}
where path $\gamma$ starts from ${\bf Y}_0=(y_0,p_{y_0})$ 
on lead $m_0$ and tunnels into the cavity,
it then bounces inside the cavity, 
finally it transmits through barrier $m$ to end at 
${\bf Y}=(y,p_y)$ in lead $m$.
We define $n_\gamma (m')$ as 
the number of times that the path $\gamma$ reflects
off the tunnel-barrier on lead $m'$ before escaping. 
The differential $(\rmd p_{y_0}/\rmd y)_\gamma $ is the
rate of change of initial momentum, $p_{y_0}$, with final position, $y$,
for an {\it unchanged set} of transmissions and reflections at the barriers.

\begin{figure*}
\begin{center}
\hspace{-1cm}
\resizebox{16cm}{!}{\includegraphics{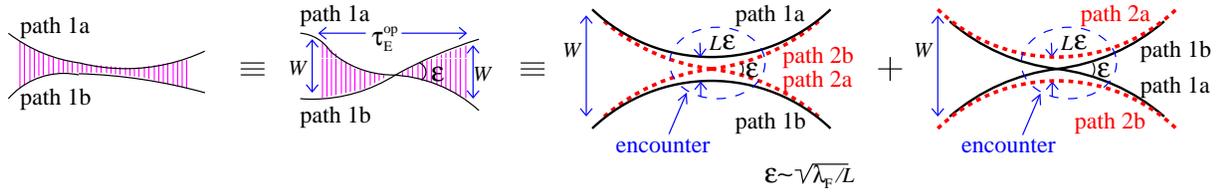}}
\caption{\label{fig:encounter}
In all figures of contributions in this article we mark the
regions where paths are correlated with vertical cross-hatching.
The encounter is at the centre of those regions.
The diagonal approximation is good everywhere except at the encounter,
thus for simplicity we show only two of the four paths (or path segments)
in all other figures in this article.  Here we show what
the cross-hatched regions in the other figures mean in terms
of all four paths that enter them.
In the cross-hatched region the two paths marked by solid lines
are paired with each other. This region is macroscopic in size,
the length of the region in time is $\tEo$ (which we assume to be
of similar magnitude to the classical timescales such as 
$\tDone,\tDtwo$), and the distance between the paths at either end of 
this region is of order the lead width.   
In contrast the encounter (where both paths marked by dashed lines swap 
from one path marked by a solid line to the other) is 
microscopic in size,  the distance between the non-crossing paths is
typically $(\lambda_{\rm F}L)^{1/2}$ which goes to zero in the classical
limit, the time which the encounter takes is a few times the inverse of the
Lyapunov exponent.
}
\end{center}
\end{figure*}

\subsection{The Landauer-B\"uttiker formula and the Drude conductance.}

Inserting Eq.~(\ref{semicl-tr}) into the Landauer-B\"uttiker
formula for the conductance 
\begin{eqnarray}
g_{mm_0}={\rm Tr} [{\mathbb S}_{mm_0}^\dagger {\mathbb S}_{mm_0}],
\end{eqnarray} 
one gets a double sum over  paths, 
$\gamma1$ and $\gamma2$ and over lead modes, $|n\rangle$ and  $|m\rangle$.
We make a semiclassical approximation (see Appendix \ref{appendix:delta_hbar}) 
that
$\sum_n\langle y'|n\rangle\langle n|y\rangle \simeq \delta (y'-y)$.
The conductance is then given by a double sum over paths 
which both go from $y_0$ on lead $m_0$ to $y$ on lead $m$,
\begin{eqnarray}
g_{mm_0}
&=& 
(2\pi \hbar)^{-1}
\!\int_{\rm L} \! \! \rmd y_0 \int_{\rm R} \! \rmd y   
\sum_{\gamma1,\gamma2} 
A_{\gamma1}A_{\gamma2}^*
\e^{\rmi\delta S/\hbar} . \quad
\label{eq:conductance}
\end{eqnarray}
where the action difference, $\delta S = S_{\gamma1}-S_{\gamma2}$.

We averaged the conductance over the energy or the cavity shape. 
For most $\{\gamma1,\gamma2\}$ 
the phase of a given contribution, $\delta S/\hbar$,
will oscillate
wildly with these variations, so the contribution 
averages to zero. 
The only contributions that will survive are those 
in which $\gamma1$ and $\gamma2$ are correlated
in such a manner that $\de S$ remains fixed 
when we vary the energy or the cavity shape.
The most obvious such contributions\cite{Berry-diag} are diagonal ones
 ($\gamma1=\gamma2$),
and we now show that they give the Drude conductance.

For these diagonal contributions, we note that 
$(\rmd p_{y_0}/\rmd y)_\gamma = p_{\rm F} \cos \theta_0
(\rmd \theta_0 /\rmd y)_\gamma$ and then use
Eq.~(\ref{eq:path-sum-as-integral}) with 
$\rho_{m'}=|t_{m'}|^2= 1-|r_{m'}|^2$,
to write the sum over
all  paths $\gamma$ from  $y_0$ to $y$ as
\begin{eqnarray}
\sum_\gamma \!
|A_\gamma|^2 \;
\! [\cdots]_\gamma 
\!\! &=& \!\!
\int_0^\infty 
\! \rmd t \; \int_{-\pi/2}^{\pi/2} \rmd \theta_0 \; \rmd \theta \; 
\nonumber \\
&& \times \; P({\bf Y},{\bf Y}_0;t) 
\; [\cdots]_{{\bf Y}_0}.
\label{eq:gamma-sum-to-Pintegral}
\end{eqnarray}
where for compactness we define
$P({\bf Y},{\bf Y}_0;t) = p_{\rm F}\cos \theta_0
\times {\tilde P}({\bf Y},{\bf Y}_0;t)$, with $p_{\rm F}\cos \theta_0$
being the initial momentum along the injection lead. 
Using Eq.~(\ref{eq:average-trans-prob}) we see that
\begin{eqnarray}
\langle P({\bf Y},{\bf Y}_0;t) \rangle = \frac{\rho_{m_0}\rho_m
p_{\rm F} \cos \theta_0 \cos \theta }
{2 (\sum_{m'} \rho_{m'} W_{m'}) \tDone} \; 
\exp[-t/\tDone] \, .
\label{eq:average-P}
\end{eqnarray}
Using Eqs.~(\ref{eq:gamma-sum-to-Pintegral}) and (\ref{eq:average-P}) 
one find the Drude conductance,
\begin{eqnarray}
g^{\rm D}_{mm_0} &=& (1-\rho_{m_0})N_{m_0} \de_{mm_0} 
+ {\rho_{m_0} N_{m_0} \rho_m N_m \over \sum_{m'}\rho_{m'} N_{m'} } \, .
\label{eq:Drude}
\end{eqnarray}
The first term above is for reflection ($\de_{mm_0}$ is a 
Kronecker delta-function), it represents 
contributions which are reflected off the barrier on the injection lead
without ever entering the cavity.

\section{Weak-localization correction to conductance 
for a chaotic system with tunnel-barriers.}
\label{sect:weak-loc}

\begin{figure}
\begin{center}
\resizebox{7.0cm}{!}{\includegraphics{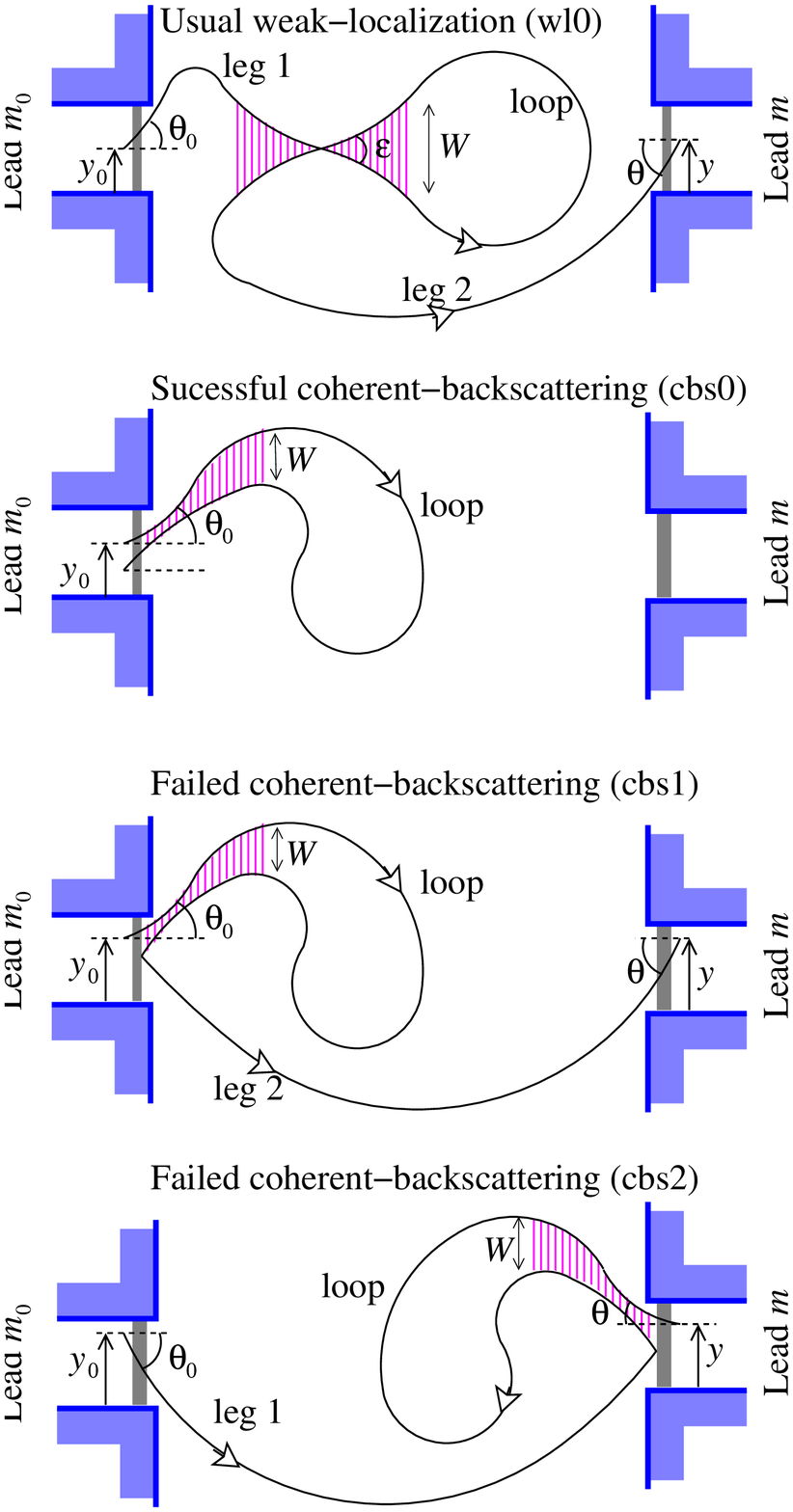}}
\caption{\label{fig:wl} 
Weak-localization contributions to conductance (contributions to 
conductance of order $N^0$) when there are tunnel-barriers
on the leads.
All contributions except the second 
are for any $m_0$ and $m$, 
while the second (cbs0) is only for $m=m_0$.
The vertical cross-hatching indicate the regions in which the 
paths are correlated with each other, so their survival probability is given by
Eq.~(\ref{eq:P2}) not Eq.~(\ref{eq:P1}).
The encounter occurs at the centre of the correlated region,
as shown in Fig.~\ref{fig:encounter}.
Details of the correlated regions for successful and failed 
coherent-backscattering are 
shown in Fig.~\ref{fig:cbs}.
}
\end{center}
\end{figure}

We identify four contributions to the weak-localization
correction to the conductance of a system coupled to leads via 
tunnel-barriers, they are shown in Fig.~\ref{fig:wl}.

\subsection{Usual weak-localization.}
\label{sect:wl0}

Here we consider the usual contribution to weak localization (wl0),
the first contribution shown in Fig.~\ref{fig:wl}.
We call it the {\it usual} contribution because it is 
the only weak-localization contribution to transmission ($m\neq m_0$) 
when the tunnel-barriers are absent ($\rho_m=1$ for all $m$).
In the usual contribution, the paths are paired almost everywhere 
except in the vicinity of an encounter\cite{Ric02}. 
At an encounter, shown in detail in Fig.~\ref{fig:encounter},
one of the paths intersects itself,
while the other one avoids the crossing.
Ref.~[\onlinecite{Sieber01}] showed that 
every self-intersecting path with a 
small crossing angle $\epsilon$ has a partner which avoids the crossing.
As a result the two paths travel along the
loop that they form in opposite directions. 
However the two paths are always close enough to each other
that they have the same stability, hence
$\sum_{\gamma1,\gamma2} A_{\gamma1}A_{\gamma2}^*
\rightarrow \sum_{\gamma1} |A_{\gamma1}|^2$.
To evaluate the usual weak-localization contribution (wl0), we 
perform a calculation similar to Ref.~[\onlinecite{jw2005}];
this broadly follows Refs.~[\onlinecite{Ric02,Ada03}] while adding 
the crucial fact that close to the encounter (the cross-hatched region in
Fig.~\ref{fig:wl}) 
the paths have a different survival probability from 
elsewhere. 
For a system with the tunnel-barriers the
survival probability is given by Eq.~(\ref{eq:P2})
inside the cross-hatched region, 
and  given by Eq.~(\ref{eq:P1}) elsewhere.

As in Ref.~[\onlinecite{jw2005}] we only consider those
contributions where the paths are uncorrelated when they hit the lead
(by which we mean a collision with a lead where at least one pair of paths 
tunnels out of the system).
Those contributions in which the paths are paired when 
one of them tunnels out of the system are included in the
coherent-backscattering contributions.
Uncorrelated escape requires a minimal time
$T_W(\eps)/2$ between encounter and escape, where \cite{caveat2}
\begin{eqnarray}
T_W(\eps) = \lambda^{-1} \ln [\eps^{-2} (W/L)^2],
\label{eq:T_W}
\end{eqnarray}  
for a small crossing angle $\eps$ (see Fig.~\ref{fig:encounter}),
in a system with Lypunov exponent $\lambda$, 
system size $L$, and lead width $W$.
This is because this is the time for 
the perpendicular distance between the paths 
to become larger than the width of the leads. 
Only then will the two paths
escape in an uncorrelated manner, typically at completely different
times, with completely different momenta (and possibly through 
different leads). 

To calculate the contribution of the sum over all paths of the form  
sketched in Fig.~\ref{fig:wl}, we note that
the action difference, $\de S$, is\cite{Sieber01,Ric02} 
\begin{eqnarray}
\delta S_{\rm wl}= E_{\rm F}\eps^2/\lambda .
\label{eq:deltaS_wl0}
\end{eqnarray}
This assumes the dynamics in the vacinity of the encounter
are time-reversal symmetric, this is the case for any applied magnetic
field which is weak enough not to affect the classical dynamics. 
We write the probability to go from ${\bf Y}_0$ to 
${\bf Y}$ in time $t$, in terms of the product of two probabilities.
The first is probability to go from
        ${\bf Y}_0$ to a point on the energy surface inside the cavity 
        ${\bf R}_1 = ({\bf r_1},\phi_1)$ (where $\phi_1$ defines 
        the direction of the momentum) in time $t_1$.
The second is the probability to go 
        from ${\bf R}_1$ to ${\bf Y}$ in time $t-t_1$. 
When one integrates this product over 
${\bf R}_1$ on the energy surface ${\cal C}$, one gets the
probability to go from ${\bf Y}_0$ to ${\bf Y}$ in time $t$.
        Thus we can write the quantity $P$, introduced above, as
\begin{eqnarray}
P({\bf Y},{\bf Y}_0;t)
\!&=& \!\!
\int_{\cal C} \rmd {\bf R}_2 \rmd {\bf R}_1
\tilde{P}({\bf Y},{\bf R}_2;t-t_2)
\nonumber \\
& \times &
\tilde{P}({\bf R}_2,{\bf R}_1;t_2-t_1)
P({\bf R}_1,{\bf Y}_0;t_1), \qquad
\end{eqnarray} 
         where $\tilde{P}({\bf R}_2,{\bf R}_1;t)$ is the probability 
         density to go from ${\bf R}_1$ to ${\bf R}_2$ in time $t$, 
         but $P({\bf R_1},{\bf Y}_0;t)$ is a probability density multiplied 
         by the injection momentum, $p_{\rm F} \cos \theta_0$.

Since we are only interested in paths that have an intersection as small crossing angle, $\eps$,
we can restrict the probabilities inside the
integral to such paths by defining ${\bf R}_1$ ( ${\bf R}_2$)
as the phase-space position for the first (second) visit to the crossing 
occurring at time $t_1$ ($t_2$). 
We can then write $\rmd {\bf R}_2 = v_{\rm F}^2 
\sin \eps \rmd t_1 \rmd t_2 \rmd \eps$
and set ${\bf R}_2\equiv({\bf r}_2,\phi_2)=({\bf r}_1,\phi_1\pm \eps)$.
Now we note that the loop cannot close before the path segments have diverged
to a distance of order the system size $L$.
When closer than this the two paths leaving an encounter
have hyperbolic relative dynamics, 
and the probability of forming a loop is zero.
Hence the duration of the loop must exceed
\begin{eqnarray}
T_L(\eps) = \lambda^{-1} \ln [\eps^{-2}],
\label{eq:T_L}
\end{eqnarray}
This means that the probability that a path starting at ${\bf Y}_0$
crosses itself  
at an angle $\pm\eps$
and then goes to lead $m$, multiplied by its injection momentum 
$p_{\rm F}\cos \theta_0$, is 
\begin{eqnarray}
I_m({\bf Y}_0,\eps)
\! &=& \! 
2v_{\rm F}^2 \; \sin \eps \;
\int_{T_L+T_W}^\infty \rmd t 
\int_{T_L+T_W/2}^{t-T_W/2} \! \rmd t_2
\int_{T_W/2}^{t_2-T_L} \! \rmd t_1
\nonumber\\
& & \times 
\int_m \rmd {\bf Y} \int_{\cal C} \rmd {\bf R}_1 \;
\tilde{P}({\bf Y},{\bf R}_2;t-t_2) 
\nonumber \\[2mm]
& & \times \,\tilde{P}({\bf R}_2,{\bf R}_1;t_2-t_1) \; P({\bf R}_1,{\bf Y}_0;t_1)  \, ,
\quad \quad 
\label{eq:I}
\end{eqnarray}
where the ${\bf Y}$-integral is over the cross-section of the $m$th lead,
and $T_W, T_L$ are shorthand for $T_W(\eps),T_L(\eps)$.

To get $g^{\rm wl0}$
we sum only contributions where $\gamma1$ crosses itself,
we then take twice the real part of this result to include the contributions
where $\gamma1$ avoids the crossing (and hence $\gamma2$ crosses itself).
Thus 
\begin{eqnarray}\label{gwl0}
g^{\rm wl0}_{mm_0}
&=& 
(\pi \hbar)^{-1}  
\int_{m_0} \rmd {\bf Y}_0 \int_0^\infty \rmd \eps 
{\rm Re}\big[\e^{\rmi \de S_{\rm wl}/\hbar}\big] 
\big\langle I_m({\bf Y}_0,\eps) \big\rangle .
\nonumber \\
\end{eqnarray}
To average $I_m({\bf Y}_0,\eps)$, we have to consider the average behaviour
of the $P$s.
Within $T_W(\eps)/2$ of the crossing the two legs of a
self-intersecting path have a joint survival probability given 
by Eq.~(\ref{eq:P2}), since they are
exploring the same region of phase-space.
Elsewhere the paths' survival probability is given by 
Eq.~(\ref{eq:P1}), because there they 
are exploring different region of phase space.
As a result the survival probability of a self intersecting path of length
$t> T_W+T_L$ is $\exp[-(t-2T_W)/\tDone -T_W/\tDtwo]$.
Since $\tDtwo^{-1} \leq 2\tDone^{-1}$, 
self-intersecting paths 
have an enhanced survival probability compared to non-crossing paths
of the same length. When the tunnel-barriers are absent, we have 
$\tDtwo=\tDone$ and the survival probability is  
$\exp[-(t-T_W)/\tDone]$, as in Refs.~[\onlinecite{jw2005,Brouwer-ucf}]
and as first noted in Ref.~[\onlinecite{Brouwer-quasi-wl}].
Outside the correlated region, the legs can escape
independently at any time through either lead. 
It is natural to assume that the probability density for the path to go
to a given point in phase-space is uniform.  
In this case, 
the probability density for leg 1 (including the tunnelling into 
the cavity from lead $m_0$) gives
\begin{eqnarray}
& & \hskip -5mm \langle P({\bf R}_1,{\bf Y}_0;t_1)\rangle 
\\
&=& \rhomo p_{\rm F}\cos \theta_0
{\exp [-(t_1-T_W/2)/\tDone -T_W/(4\tDtwo)] \over 2\pi A} ,
\nonumber 
\end{eqnarray}
the loop's probability density is
\begin{eqnarray}
& & \hskip -5mm \langle \tilde{P}({\bf R}_2,{\bf R}_1;t_2-t_1)\rangle 
\\
&=& {\exp [-[t_2-t_1-T_W/2]/\tDone -T_W/(2\tDtwo) ] \over 2\pi A} ,
\nonumber
\end{eqnarray}
and the probability density for leg 2 (including tunnelling into lead $m$) is
\begin{eqnarray}
& &  \hskip -5mm \langle \tilde{P}({\bf Y},{\bf R}_2;t-t_2)\rangle 
\\
&=& \rhom {\cos \theta
\exp [-(t-t_2-T_W/2)/\tDone -T_W/(4\tDtwo)] \over 
2\tDone \sum_{m'} \rho_{m'}W_{m'}  }.
\nonumber
\end{eqnarray}
Note that all the above probabilities are {\it conditional} 
on the fact the path has an encounter, so that the pair of path segments 
within $T_W/2$ of that encounter have a joint
survival probability given by Eq.~(\ref{eq:P2})
(for convenience we divide that joint survival
probability equally between the path segments).
Thus we find that
\begin{eqnarray}
& & \hskip -5mm
\langle \tilde{P}({\bf Y},{\bf R}_2;t-t_2) \; 
\tilde{P}({\bf R}_2,{\bf R}_1;t_2-t_1) \; 
P({\bf R}_1,{\bf Y}_0;t_1) \rangle  
\nonumber \\
&=&\frac{\rhomo\rhom}{(2 \pi A)^2} \; 
\frac{p_{\rm F} \cos \theta \cos \theta_0}
{2\tDone \sum_{m'} \rho_{m'}W_{m'}  }  
\nonumber \\
& & \times \exp[-(t-2T_W)/\tDone - T_W/\tDtwo] , \quad \quad  
\end{eqnarray}
so that $\langle I_m( {\bf Y}_0,\eps) \rangle$ becomes 
\begin{eqnarray}
\big\langle I_m({\bf Y}_0,\eps) \big\rangle
&=&{\rhomo\rhom (v_{\rm F} \tDone)^2 \over \pi A} \; 
{N_{\rm m} p_{\rm F} \sin \epsilon \cos \theta_0 
\over  \sum_{m'} \rho_{m'}N_{m'} }
\nonumber \\
& & \times \exp[-T_W/\tDtwo -(T_L-T_W)/\tDone]. 
\nonumber \\
\label{eq:average-I}
\end{eqnarray}
We insert this into Eq.~(\ref{gwl0}). The integral over $\epsilon$ 
is dominated by contributions with $\epsilon \ll 1$, so that we
write $\sin \eps \simeq \eps$ and push the upper bound for the
$\epsilon$-integration to infinity. Then the integral over $\eps$ 
is\cite{Ada03}, 
\begin{eqnarray}
& & \hskip -8mm
{\rm Re}
\left[\int_0^\infty \eps \, \rmd \eps  
\left({L\eps/ W}\right)^{2/(\lambda \tDtwo)} 
\exp \left[{\rmi \EF \eps^2\over \lambda\hbar } \right] \right]
\nonumber \\
&=&
{\lambda \hbar \over 2 \EF} 
{\rm Re}\!\left[\rmi^{1+(\lambda\tDtwo)^{-1}} \right]
\Gamma[1\!+\!(\lambda\tDtwo)^{-1}]
\left(\! {\lambda \hbar L^2 \over \EF W^2}\! \right)^{(\lambda\tDtwo)^{-1}}
\nonumber \\
&=&
-{\pi \hbar \over 2mv_{\rm F}^2 \tDtwo} \exp[-\tEo/\tDtwo] 
+ {\cal O}[(\lambda\tDtwo)^{-1}],
\label{eq:eps-integral}
\end{eqnarray}
where to get the second line we expanded to leading order in
$(\lambda\tDtwo)^{-1}$, in this situation the
Euler Gamma-function, $\Gamma[1+(\lambda\tDtwo)^{-1}]\simeq1$.
Note that when we neglect all ${\cal O}[(\lambda\tDtwo)^{-1}]$ corrections
this must include all order-one terms in the logarithm of the Ehrenfest time,
since they can only lead to ${\cal O}[(\lambda\tDtwo)^{-1}]$ corrections
to the above expression.  Thus since $\lambda \sim v_{\rm F}/L$,
we are justified in writing
$\tEo= \lambda^{-1} \ln[\EF W^2/(\lambda \hbar L^2) ]$.

Note that the second term in the exponent of Eq.~(\ref{eq:average-I})
is independent of $\eps$, thus only $\tDtwo$ enters the $\eps$-integral.
The ${\bf Y}_0$-integral generates a factor of $2 W_{m_0}$. 
We can write $N_m=(\pi \hbar)^{-1}p_{\rm F}W_m$
and $(\sum_{m'} \rho_{m'}N_{m'})^{-1}
= (mA)^{-1}\hbar \tDone$.  Thus the usual weak localization 
contribution is 
\begin{eqnarray}
g^{\rm wl0}_{mm_0} 
&=& -{\rhom \rhomo \Nm \Nmo \over [\sum_{m'}\rho_{m'}N_{m'}]^2}  
\,{\tDone \over \tDtwo}
\nonumber \\
& & \times 
\exp [-\tEo/\tDtwo - (\tEc-\tEo)/\tDone], 
\label{eq:wl0}
\end{eqnarray}
where $\tDone$,$\tDtwo$
are given in Eqs.~(\ref{eq:tDone},\ref{eq:tDtwo}).
When the tunnel-barriers are transparent ($\rho_m=1$ for all $m$),
this is the only contribution to weak-localization for $m\neq m_0$;
in this case $\tDtwo=\tDone$ and we get the 
expected result\cite{jw2005,Brouwer-quasi-wl,Brouwer-ucf}.

Note that in the universal limit ($\tEo\to 0$)
this contribution to conductance behaves like the naive argument
given in Section~\ref{sect:intro} would predict.  
It goes like the 
weak-localization contribution to conductance for a cavity with
{\it no} tunnel-barriers and each lead having $N_i'$ modes where
$N_i'=\rho_iN_i$,
(with $\tDone/\tDtwo$ being just a numerical  prefactor with
$1 \leq \tDone/\tDtwo  \leq 2$).  
Thus in the limit of opaque barriers ($\rho_m \to 0$ for all $m$),
this contribution does {\it not} vanish for fixed Drude conductance
($\rho_mN_m$ remains constant for all $m$). 
Thus the origin of the suppression of 
weak-localization in the opaque limit is elsewhere.

\subsection{Successful coherent-backscattering.}
\label{sect:cbs0}

\begin{figure}
\begin{center}
\hspace{-1cm}
\resizebox{8cm}{!}{\includegraphics{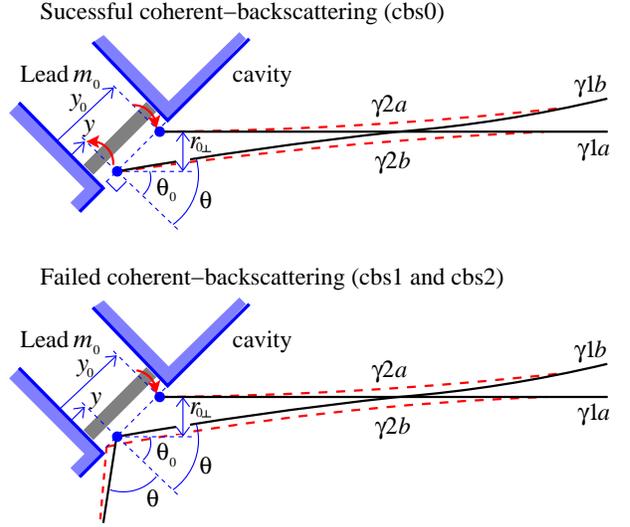}}
\caption{\label{fig:cbs} 
Details of the paths that contribute to successful and failed 
coherent-backscattering (cbs0, cbs1 and cbs2).
Path $\gamma 1$ (solid black line)
start on the cross-section of lead $m_0$ at position
$y_0$ with momentum angle $\theta_0$ and
returns at $y$ with momentum angle $\theta$.
For the successful contribution it then tunnels into the lead,
for the failed contribution it is reflected off the barrier and
remains in the chaotic system.
All paths are drawn in the 
basis parallel and perpendicular to $\gamma1$ at injection, 
the initial position and momentum of path $\gamma1$ at exit are
$r_{0\perp} = (y_0-y)\cos \theta_0$, $r_{0\pll} = (y_0-y)\sin \theta_0$  
and $p_{0\perp} \simeq -p_{\rm F} (\theta-\theta_0)$.}
\end{center}
\end{figure}

The coherent-backscattering contribution to transport has 
long been considered in trajectory-based semiclassics\cite{Bar93,Ric02}.
However until recently the approximations were a little simplistic, and failed
to capture the full nature of this contribution 
(in particular its Ehrenfest time dependence).  This was addressed
in Ref.~[\onlinecite{Brouwer-cbs,jw2005}] in the absence of tunnel-barriers.
Here we perform a calculation similar to Ref.~[\onlinecite{jw2005}],
while adding tunnel-barriers on the leads.
The coherent-backscattering contributions in the presence of these barriers 
are shown in Fig.~\ref{fig:wl},
with Fig.~\ref{fig:cbs} showing the paths in the correlated region 
in more detail. 
We treat these contributions separately
from those in the previous section because
injection and exit positions and momenta are correlated.
These contributions are exactly those that were ignored in the previous section,
paths where one/both legs escape within $T_W/2$ of the encounter.
However it is convenient to parameterize these contributions 
in terms of $(r_{0\perp},p_{0\perp})$ instead of $(\eps, t_1)$.

Here we consider the successful coherent-backscattering (cbs0),
the second contribution shown in Fig.~\ref{fig:wl}.
This {\it successful} contribution gives the usual 
coherent-backscattering 
when the tunnel-barriers are absent ($\rho_m=1$ for all $m$).
From Ref.~[\onlinecite{jw2005}], we have that 
\begin{eqnarray}
S_{2b}-S_{1a} &=& \pF r_{0\pll} + \half m\lambda r_{0\perp}^2 ,
\label{eq:cbs0-correlated-action-diff1}
\\
S_{2a}-S_{1b} &=& -\pF r_{0\pll} 
+ p_{0\perp}r_{0\perp} +\half m\lambda r_{0\perp}^2 , 
\label{eq:cbs0-correlated-action-diff2}
\end{eqnarray}
where $(r_{0\perp},p_{0\perp})$ are the position and momentum
of path segment $\gamma1b$ relative to path segment $\gamma1a$, 
shown in Fig.~\ref{fig:cbs}.
Thus 
the total action difference between these two paths is
\begin{eqnarray}
\de S_{\rm cbs0} 
&=& (p_{0\perp}+m\lambda r_{0\perp})r_{0\perp}\,.
\label{eq:deltaS_cbs}
\end{eqnarray}
The successful coherent-backscattering contribution to the reflection is
\begin{eqnarray}
g^{\rm cbs0}_{mm_0} 
&=& \de_{mm_0} (2\pi \hbar)^{-2}
\!\int_{\rm m_0} \!\! \rmd {\bf Y}_0  \rmd {\bf Y}
\int_0^\infty \!\! \rmd t 
\nonumber \\ 
& & \qquad \times
\langle P({\bf Y},{\bf Y}_0;t)  \rangle \;{\rm Re}\big[
\e^{\rmi \de S_{\rm cbs0}/\hbar}\big] ,
\label{eq:R_cbs}
\end{eqnarray}
where $\de_{mm_0}$ is a Kronecker $\de$-function.

To perform the average 
we define $T'_{W}(r_{0\perp}, p_{0\perp})$ 
and $T'_{L}(r_{0\perp}, p_{0\perp})$ 
as the time between touching the tunnel barrier 
and the perpendicular distance between
$\gamma1a$ and $\gamma1b$ becoming $W$ and $L$, respectively.
For times less than $T'_{W}(r_{0\perp}, p_{0\perp})$, 
the path segments ($\gamma1a$ and $\gamma1b$) 
are paired thus their joint survival probability is given by 
Eq.~(\ref{eq:P2}). 
For times longer than this
the path segments escape independently, so each has a survival probability
given by Eq.~(\ref{eq:P1}).
For $g^{\rm cbs0}_{mm_0}$ we consider only those paths
that form a closed loop, however they cannot close until the two 
path segments are of order $L$ apart. 
This means that the $t$-integral in eq.~(\ref{eq:R_cbs}) 
must have a lower cut-off at 
$2T'_L(r_{0\perp}, p_{0\perp})$,
hence we have 
\begin{eqnarray}
& & \hskip -8mm \int_{m_0} \! \rmd {\bf Y}
\int_{2T'_L}^\infty \rmd t 
\langle P({\bf Y},{\bf Y}_0;t)\rangle 
\nonumber \\ 
&=& \rhomo^2 
p_{\rm F} \cos \theta_0 \; \frac{N_{m_0}}{\sum_{m'} \rho_{m'}N_{m'} } 
\nonumber \\
& & \times
\exp[-T'_W/\tDtwo -2(T'_L-T'_W)/\tDone] ,\  
\, \qquad
\end{eqnarray}
where $T'_{L,W}$ are shorthand for $T'_{L,W}(r_{0\perp}, p_{0\perp})$.
For small $(p_{0\perp}+ m\lambda r_{0\perp})$ we estimate \cite{typo}
\begin{eqnarray}
T'_W(r_{0\perp}, p_{0\perp}) 
&\simeq& 
\lambda^{-1} \ln \left[
{ m\lambda W \over |p_{0\perp}+ m\lambda r_{0\perp}| }\right] , \qquad 
\label{eq:T'_W}
\\
T'_L(r_{0\perp}, p_{0\perp}) 
&\simeq& 
\lambda^{-1} \ln \left[
{ m\lambda L \over |p_{0\perp}+ m\lambda r_{0\perp}| }\right] . \qquad 
\label{eq:T'_L}
\end{eqnarray}
Thus $(T'_L-T'_W)$ is independent of $(r_{0\perp}, p_{0\perp})$,
indeed $2(T'_L-T'_W)= (\tEc-\tEo)$.
We substitute the above expressions into $g^{\rm cbs0}_{mm_0}$,
write\cite{Bar93} 
$p_{\rm F} \cos\theta_0 \rmd {\bf Y}_0 =  
\rmd y_0\rmd (p_{\rm F} \sin\theta_0 )
= \rmd r_{0\perp} \rmd p_{0\perp}$, and then make the substitution
$\tilde{p}_0=p_{0\perp}+m\lambda r_{0\perp}$.
We evaluate the $r_{0\perp}-$integral over a range of order $W_{m_0}$.
Then, taking the limits on the resulting $\tilde{p}_0$-integral 
to infinity, it takes the form
\begin{eqnarray}
& & \hskip -8mm \int_{-\infty}^\infty \rmd \tilde{p}_0 
{2\hbar \sin(\tilde{p}_0 W/\hbar)\over \tilde{p}_0}
 \left| \tilde{p}_0 \over m\lambda W \right|^{1/(\lambda\tDtwo)}
\nonumber \\
&=& { 4\hbar \over (m\lambda W )^{1/(\lambda\tDtwo)}} 
{\rm Im} \left[ \int_0^\infty \rmd \tilde{p}_0 \, 
\tilde{p}_0^{-1+ 1/(\lambda\tDtwo)} 
\exp[\rmi\tilde{p}_0 W/\hbar] \right]
\nonumber \\
&=& 2\pi \hbar \left(\hbar \over m\lambda W^2 \right)^{1/(\lambda\tDtwo)} .
\label{eq:tilde_p-int1}
\end{eqnarray}
To evaluate the integral we wrote it in terms of an 
Euler $\Gamma$-function.
Note that $[\hbar/(m\lambda W^2)]^{1/(\lambda\tDtwo)}
\simeq \exp[-\tEo/\tDtwo]$ once we have dropped ${\cal O}(1)$-terms 
inside logarithm. The result is that
\begin{eqnarray}
g^{\rm cbs0}_{mm_0} 
&=& \de_{mm_0} {\rhomo^2 \Nmo \over \sum_{m'}\rho_{m'}N_{m'} } 
\nonumber \\
& & \times 
\exp \Big[-\tEo/\tDtwo - (\tEc-\tEo)/\tDone] . \quad 
\label{eq:cbs0}
\end{eqnarray}
Note that the successful coherent-backscattering contribution has 
exactly the same exponential dependence on $\tEo,\tEc$
as the successful weak-localization. 
This must be the case if the theory is to preserve unitarity
(conserve current), however since wl0 and cbs0 were calculated 
separately using different methods, this acts as a check
on our algebra.

\subsection{Failed coherent-backscattering.}
\label{sect:cbs1}

Here we consider the failed coherent-backscattering (cbs1 and cbs2),
see Fig.~\ref{fig:wl}c,d.
We call them {\it failed} contributions because 
they do not exist in the absence of the tunnel-barriers.
They are part of the coherent-backscattering peak for the chaotic system,
however they are reflected back into the cavity by the tunnel-barriers. 
They can then contribute to either transmission or reflection.

The action difference for coherent-backscattering paths 
that reflect at the barrier 
is slightly different from that for the paths that transmit.
This is because the path segments $\gamma1b$ and $\gamma2b$
Converge at infinity rather than at the lead (see Fig.~\ref{fig:cbs}).
We split path segment $\gamma1b$ into 
$\gamma1b'$ before the failure to tunnel
and $\gamma1b''$ after the failure to tunnel.
We do the same for path segment $\gamma2b$.
Then the action difference is the sum of three terms.
The first two come from before the failed tunnelling;
they are
\begin{eqnarray}
S_{2b'}-S_{1a} &=& \pF r_{0\pll}
+ \half m\lambda \big[ r_{0\perp}/2-p_{0\perp}/(2m\lambda)]^2 \, . \quad
\end{eqnarray}
and $(S_{2a}-S_{1b'})$ given by 
eq.~(\ref{eq:cbs0-correlated-action-diff2}).
The third term is the action difference of path segments 
after the failure to tunnel,
\begin{eqnarray}
S_{2b''}-S_{1b''} &=& - p_{0\perp}\big[r_{0\perp}+p_{0\perp}/(m\lambda)\big]
\nonumber \\
& & + \half m\lambda \big[ r_{0\perp}+p_{0\perp}/(m\lambda)]^2 \, . \quad
\end{eqnarray}
Hence the total action difference between the two paths
\begin{eqnarray}
\de S_{\rm cbs1} = \de S_{\rm cbs0} + \tilde p_0^2/(4m\lambda) 
\label{eq:deltaS_cbs-indir}
\end{eqnarray}
where we write $\tilde p_0 = p_{0\perp} + m\lambda r_{0\perp}$
and hence $\de S_{\rm cbs0} = r_{0\perp}\tilde p_0$,
as in Section \ref{sect:cbs0}.
As before we evaluate the integral over $r_{0\perp}$ first,
getting an integral for $\tilde p_0$ of the form
\begin{eqnarray}
& & \hskip -8mm \int_{-\infty}^\infty \rmd \tilde{p}_0 
{2\hbar \sin(\tilde{p}_0 W/\hbar)\over \tilde{p}_0} 
\exp \left[{\rmi \tilde p_{0}^2 \over 4m\lambda\hbar }\right]
 \ \left| \tilde{p}_0 \over m\lambda W \right|^{1/(\lambda\tDtwo)}
\nonumber \\
&=& 2\pi \hbar \left(\hbar \over m\lambda W^2 \right)^{1/(\lambda\tDtwo)}.
\end{eqnarray}
To get this result we noted that $ p_{0} \lesssim \hbar/W$
in the region which dominates the integral, so we
approximated
$\exp [\rmi \tilde{p}_0^2 (4m\lambda\hbar)^{-1}]=1$,
after this the integral is identical to Eq.~(\ref{eq:tilde_p-int1}).

For the first failed backscattering contribution (cbs1) 
the bulk of the derivation is 
identical to that of the
successful coherent-backscattering (cbs0) above.
The difference here is that at the point where the path returns to the lead
it is reflected from the barrier (with probability $1-\rho_{m_0}$)
instead of being transmitted (with probability $\rho_{m_0}$).
The path then remains in the cavity, and will eventually escape through
an arbitrary lead, with the probability of escape through lead $m$ being
$[\sum_{m'}\rho_{m'} N_{m'}]^{-1}\rhom\Nm$.
Thus we can conclude that 
\begin{eqnarray}
g^{\rm cbs1}_{mm_0} 
&=& g^{\rm cbs0}_{m_0m_0} \times {1-\rho_{m_0} \over \rho_{m_0}}\,
{\rhom\Nm \over \sum_{m'}\rho_{m'}N_{m'}}  
\nonumber \\
&=& {\rhom \rhomo (1-\rhomo) \Nm \Nmo \over [\sum_{m'}\rho_{m'}N_{m'} ]^2} 
\nonumber \\
& & \times 
\exp [-\tEo/\tDtwo - (\tEc-\tEo)/\tDone] 
\label{eq:cbs1}
\end{eqnarray}

The second failed coherent-backscattering contribution (cbs2)
can be immediately written down by swapping all subscripts $m_0$ with $m$,
hence 
\begin{eqnarray}
g^{\rm cbs2}_{mm_0} 
&=& {\rhom \rhomo (1-\rhom) \Nm \Nmo \over [\sum_{m'}\rho_{m'}N_{m'} ]^2} 
\nonumber \\
& & \times 
\exp [-\tEo/\tDtwo - (\tEc-\tEo)/\tDone] 
\label{eq:cbs2}
\end{eqnarray}

For $m=m_0$, one must confirm that 
cbs1 and cbs2 are not double-counting the same contribution.
One can see they form different contributions from the fact that 
one starts with a leg (two paths identical)
and ends with a loop (two paths non-identical), 
while the other starts with the loop and ends with a leg.

\subsection{The total weak-localization correction with 
tunnel-barriers.}

We sum 
the contributions calculated in the the preceding subsections,
Eqs.~(\ref{eq:wl0},\ref{eq:cbs0},\ref{eq:cbs1},\ref{eq:cbs2}), and get
\begin{eqnarray}
g^{\rm wl}_{mm_0} 
&=& - {\rhom \rhomo \Nm \Nmo \over [\sum_{m'}\rho_{m'}N_{m'}]^2} 
{\cal A}_{mm_0}
\nonumber \\
& & \times 
\exp[ -\tEo/\tDtwo - (\tEc-\tEo)/\tDone],
\label{eq:wl-m-neq-m0}
\end{eqnarray}
where we define
\begin{eqnarray}
{\cal A}_{mm_0}
&\equiv& \rhom +\rhomo -
{\sum_{m'}\rho_{m'}^2N_{m'}\over \sum_{m'}\rho_{m'}N_{m'}} 
-{\de_{mm_0} \over\Nmo} \sum_{m'}\rho_{m'}N_{m'},  
\nonumber \\
\end{eqnarray}
with $\de_{mm_0}$ being a Kronecker $\de$-function.
Note that ${\cal A}$ goes like $(\rho \times N^0)$. Thus in the limit
$\rho \to 0$ with constant $\rho N$, we see that ${\cal A}\to 0$ and the
weak-localization contribution to conductance goes to zero.

As a check on our result we verify that $\sum_m g^{\rm wl}_{mm_0} =0$,
this  
means that we have not violated unitarity (or current conservation).

Finally we consider the special case of 
two leads (L and R) with barriers with tunnel probability $\rho_{\rm L,R}$
and number of modes $N_{\rm L,R}$, then we find that  
Eq.~(\ref{eq:wl-m-neq-m0}) reduces to
\begin{eqnarray}
\de g^{\rm wl}_{\rm LR} 
&=& - {\rho_{\rm L}^2\rho_{\rm R}^2 N_{\rm L}N_{\rm R}(N_{\rm L} +N_{\rm R}) 
\over (\rho_{\rm L} N_{\rm L} + \rho_{\rm R} N_{\rm R})^3}
\nonumber \\
& & \times 
\exp [-\tEo/\tDtwo - (\tEc-\tEo)/\tDone] 
\end{eqnarray}
with 
$\de g^{\rm wl}_{\rm LL}=\de g^{\rm wl}_{\rm RR}
= -\de g^{\rm wl}_{\rm RL}= -\de g^{\rm wl}_{\rm LR}$.

\begin{figure*}
\begin{center}
\resizebox{15cm}{!}{\includegraphics{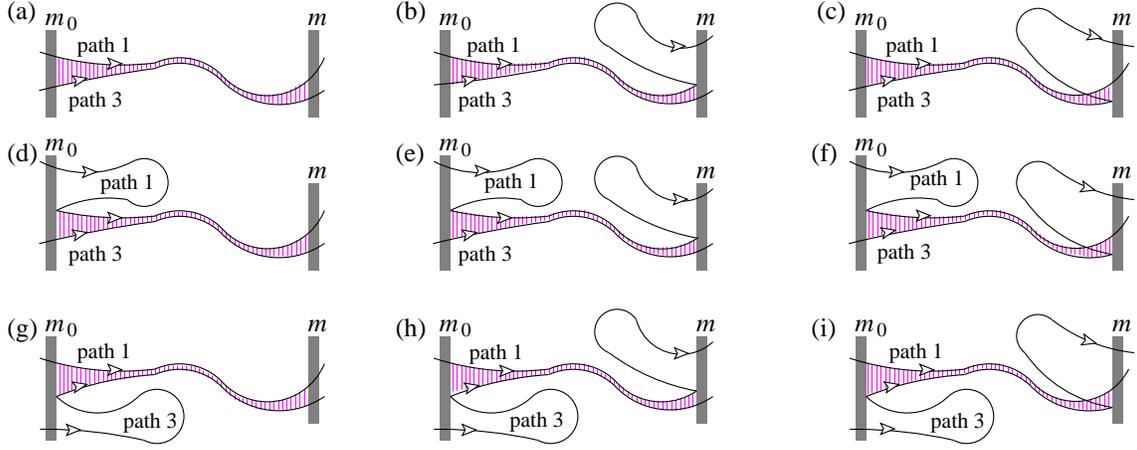}}
\caption{\label{fig:noise-trans1}
The set of contributions to 
${\rm tr}[{\mathbb S}_{mm_0}^\dagger {\mathbb S}_{mm_0}
{\mathbb S}_{mm_0}^\dagger {\mathbb S}_{mm_0}]$
which we call $D_{\rm 1a},\cdots,D_{\rm 1i}$. 
These are the contributions 
which do not vanish for infinite Ehrenfest time. 
Here we show only the tunnel-barriers on leads $m_0$ and $m$ as shaded 
rectangles, a path which crosses the barrier on lead $m$ has succeeded
in tunnelling out of the cavity into the lead.  
The contributions are made up of four classical paths,
here we show only two of the paths (1 and 3). The other two paths (2 and 4)
look the same as the paths shown, {\it except} that they cross
at the centre of the correlated region (indicated by the vertical 
crosshatching).  In the correlated region the paths have the same topology 
as shown in Fig.~\ref{fig:encounter} if we replace the labels in the manner
given in Eq.~(\ref{eq:replacements}).
Thus path 4 is paired with path 1 at lead $m_0$ but
paired with path 3 at lead $m$ (and vice-versa for path 2).
The noise in these contribution is purely due to the stochastic nature 
of scattering at the tunnel-barriers, if they were absent these
contributions would be noiseless.  
The calculation will show that 
all these contributions go like $(1-\e^{-\tEo/\tDtwo})$.
}
\end{center}
\end{figure*}

\section{Shot-noise for a chaotic system with tunnel-barriers.}
\label{sect:noise}

\begin{figure*}
\begin{center}
\hspace{-1cm}
\resizebox{14cm}{!}{\includegraphics{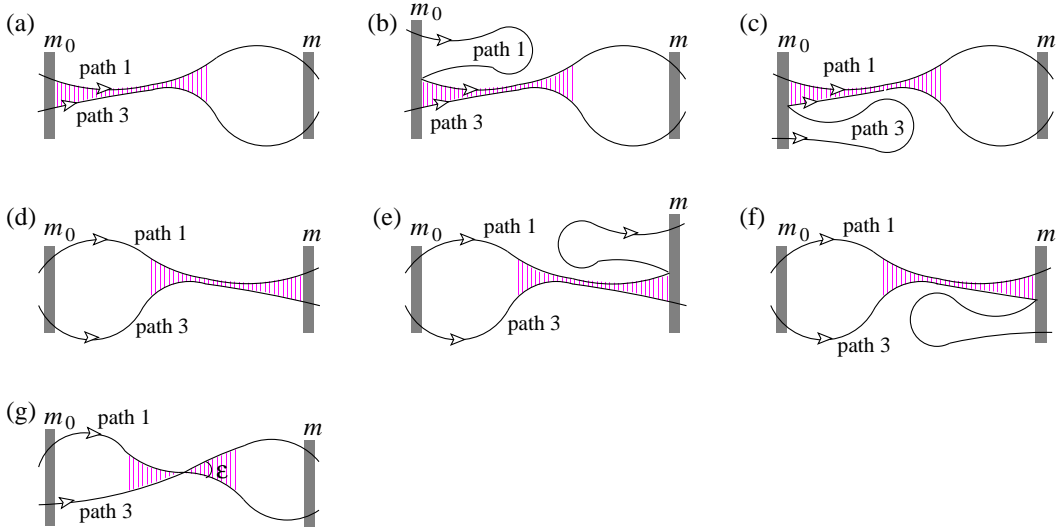}}
\caption{\label{fig:noise-trans2}
The set of contributions to 
${\rm tr}[{\mathbb S}_{mm_0}^\dagger {\mathbb S}_{mm_0}
{\mathbb S}_{mm_0}^\dagger {\mathbb S}_{mm_0}]$
which we call $D_{\rm 2a},\cdots,D_{\rm 2g}$. These are the contributions 
which vanish for infinite Ehrenfest time.
The contributions are drawn in the same manner as Fig.~\ref{fig:noise-trans1},
with only paths 1 and 3 shown. As before paths 2 and 4 are equivalent 
to 1 and 3 except that they cross at the centre of the correlated region
(indicated by vertical crosshatching).
For finite Ehrenfest times, the paths must escape the 
correlated region (vertical crosshatching) 
before either tunnels out of the cavity 
(on at least one side of the encounter). 
The calculation will show that
all these contributions go like $\e^{-\tEo/\tDtwo}$.
}
\end{center}
\end{figure*}

Now we turn to the zero-frequency shot noise
power, $S$, for quantum chaotic systems. 
This intrinsically quantum part of the fluctuations of 
a non-equilibrium electronic current
often contains information on the system that cannot be obtained through
conductance measurements. 
We give our results in terms of the Fano factor $F=S/S_{\rm p}$,
which is the ratio of $S$ to the Poissonian noise 
$S_{\rm p}=2 e \langle I \rangle$ that would be generated by a current 
flow of uncorrelated particles.
Alternatively, one can think of the Fano factor as the 
ratio of the noise to the average current,
written in convenient dimensionless units.
According to the scattering theory of
transport one has\cite{Blanter-review}
\begin{eqnarray}
F&=& 
{{\rm Tr}[{\mathbb S}_{mm_0}^\dagger {\mathbb S}_{mm_0}]
-{\rm Tr}[{\mathbb S}_{mm_0}^\dagger {\mathbb S}_{mm_0}
{\mathbb S}_{mm_0}^\dagger {\mathbb S}_{mm_0}] 
\over 
{\rm Tr}[{\mathbb S}_{mm_0}^\dagger {\mathbb S}_{mm_0}] }. \quad
\label{eq:Fano-definition}
\end{eqnarray}
We use the trajectory-based method developed in 
Ref.~[\onlinecite{wj2005-fano}] to evaluate the contributions
to this quantity to lowest order in $N^{-1}$, where $N$ is the number of 
lead modes. 
We thereby neglect all the weak-localization-type corrections to the noise
calculated in Ref.~[\onlinecite{Haake-fano}].
At this order, all contributions are listed in Figs.~\ref{fig:noise-trans1}
and \ref{fig:noise-trans2}. 
The contributions fall naturally into two classes.
The first class,
shown in Fig.~\ref{fig:noise-trans1}, 
involve classical paths (path 1 and 3) which are correlated,
and one or both paths escape {\it before} their flow under the cavity dynamics makes them become uncorrelated.
The second class, shown in Fig.~\ref{fig:noise-trans2},
involve correlated-paths which become uncorrelated under the cavity's classical
dynamics before any paths escape.
As before, {\it correlated} means 
that the paths are almost parallel and within $W$ of each other.

The denominator and the first term in the numerator of 
Eq.~(\ref{eq:Fano-definition}) are equal to the Drude conductance, 
and are hence given by
Eq.~(\ref{eq:Drude}) with $m\neq m_0$.  
Thus to find the Fano factor we must evaluate 
${\rm Tr}[{\mathbb S}_{mm_0}^\dagger {\mathbb S}_{mm_0}
{\mathbb S}_{mm_0}^\dagger {\mathbb S}_{mm_0}]$. 
As with weak-localization we approximate 
$\sum_n\langle y'|n\rangle\langle n|y\rangle \approx 
\delta (y'-y)$, then it becomes 
a sum over four paths, $\gamma1$ from $y_{01}$ to $y_1$,
$\gamma2$ from $y_{03}$ to $y_1$,
$\gamma3$ from $y_{03}$ to $y_3$ and
$\gamma4$ from $y_{01}$ to $y_3$, 
where $y_{01},y_{03}$ are on lead $m_0$ and  $y_1,y_3$
are on lead $m$.
Hence
\begin{eqnarray}\label{trt4}
& & \hskip -12mm
{\rm Tr}[{\mathbb S}_{mm_0}^\dagger 
{\mathbb S}_{mm_0}{\mathbb S}_{mm_0}^\dagger {\mathbb S}_{mm_0}]
\nonumber \\
&=& 
{1\over (2\pi \hbar)^2}
\!\int_{\rm L} \! \! 
\rmd y_{01} \rmd y_{03} \int_{\rm R} \! \rmd y_1 \rmd y_3  \nonumber \\
& & \times 
\sum_{\gamma1,\cdots \gamma4} 
A_{\gamma4}^*A_{\gamma3} A_{\gamma2}^*A_{\gamma1}
\exp [\rmi\delta S/\hbar] \,.
\end{eqnarray}
where $A_\gamma$ is given by Eq.~(\ref{eq:A}),
and $\delta S = S_{\gamma1}-S_{\gamma2}+S_{\gamma3}-S_{\gamma4}$
(we have absorbed all Maslov indices into the actions
$S_{\gamma i}$).
The dominant contributions that survive averaging over energy or cavity shape
are those for which
the fluctuations of $\delta S/\hbar$ are minimal. They are shown in 
Figs.~\ref{fig:noise-trans1},\ref{fig:noise-trans2}. 
Their paths are in pairs almost everywhere except in the
vicinity of encounters. Going through an encounter,
two of the four paths cross each other, while the other two
avoid the crossing. They remain in pairs, though the pairing switches,
e.g. from $(\gamma1;\gamma4)$ and $(\gamma2;\gamma3)$ to 
$(\gamma1;\gamma2)$ and $(\gamma3;\gamma4)$.
Paths are always close enough to their partner
that their stability is the same.
Thus, for all pairings
\begin{equation}\label{paths-pairing}
\sum_{\gamma1,...\gamma4} A_{\gamma4}^*A_{\gamma3} A_{\gamma2}^*A_{\gamma1}
\rightarrow \sum_{\gamma1,\gamma3} |A_{\gamma3}|^2 |A_{\gamma1}|^2.
\end{equation}
Then the sum over
all paths $\gamma$ from  $y_0$ to $y$ is
given by Eq.~(\ref{eq:gamma-sum-to-Pintegral})
As with weak-localization, we define 
$P({\bf Y},{\bf Y}_0;t)\de y\de \theta \de t$
as the product of the momentum along the injection lead,
$p_{\rm F}\cos \theta_0$, and the classical probability to go
from an initial position and angle
${\bf Y}_0=(y_0,\theta_0)$ to within $(\de y,\de \theta)$ of
${\bf Y}$ in a time within $\de t$ of $t$. 
We remind the reader that this probability is to go from the injection
lead to the exit lead, and this includes the tunnelling probability at each
barrier. 
We now use Eqs.~(\ref{trt4}), (\ref{paths-pairing})
and (\ref{eq:gamma-sum-to-Pintegral}) to
analyze the contributions in
Figs.~\ref{fig:noise-trans1},\ref{fig:noise-trans2}.
All contributions can be written as
\begin{eqnarray}\label{contribution}
D_i 
=
{1\over (2\pi \hbar)^2}
\!\int_{\rm L} \! \! \rmd {\bf Y}_{01} \; \rmd {\bf Y}_{03} 
\!\int_{\rm R} \! \! 
\rmd {\bf Y}_{1} \; \rmd {\bf Y}_{3} \!\int \! \! \rmd t_1 \; \rmd t_3 
\nonumber \\
\times \;
\langle P({\bf Y}_1,{\bf Y}_{01};t_1) \;  P({\bf Y}_3,{\bf Y}_{03};t_3) 
\rangle \; \exp [\rmi\delta S_{D_i}/\hbar] \,,
\end{eqnarray}
where the subscripts $1,3$ indicate paths 1 and 3 respectively. 
When evaluating Eq.~(\ref{contribution}) the joint exit probability for two
crossing paths has to be computed.

\subsection{Noise contributions 
which do not vanish for infinite Ehrenfest time.}
\label{sect:D1}

To evaluate all the contributions in Fig.~\ref{fig:noise-trans1}, we note
that the paths never become uncorrelated under the classical dynamics;
they only escape in an uncorrelated manner if one path tunnels while the 
other is reflected.
In this case the details of the
encounter are as given in Fig.~\ref{fig:cbs}, 
if we make the replacements
\begin{eqnarray} 
& & \hbox{ path segment $\gamma 1a$ $\longrightarrow$  path $\gamma 1$,} 
\nonumber \\
& & \hbox{ path segment $\gamma 2a$ $\longrightarrow$  path $\gamma 2$,} 
\nonumber \\
& & \hbox{ path segment $\gamma 1b$ $\longrightarrow$  path $\gamma 3$,}
\nonumber \\
& & \hbox{ path segment $\gamma 2b$ $\longrightarrow$  path $\gamma 4$.}
\label{eq:replacements}
\end{eqnarray}
Thus the action difference between 
the paths can be evaluated in a manner equivalent 
to coherent-backscattering.
For contributions where both paths escape as a pair, the action difference 
is given by $\de S_{D_i}= \de S_{\rm cbs0}$ where 
$\de S_{\rm cbs0}$ is given in Eq.~(\ref{eq:deltaS_cbs}). 
For contributions where the pair is broken by one of the paths 
escaping, the action
difference is given by $\de S_{D_i}=\de S_{\rm cbs1}$ with $\de S_{\rm cbs1}$ 
given in Eq.~(\ref{eq:deltaS_cbs-indir}).
Here, as in Section~\ref{sect:cbs1},  
the difference between $\de S_{\rm cbs1}$ and  $\de S_{\rm cbs0}$
has no effect on the final result, so one could just  
use $\de S_{\rm cbs0}$ for all contributions.

For the contribution in Fig.~\ref{fig:noise-trans1}a,
the paths paired at lead $m_0$ remain paired
at lead $m$, thus the length of the paired paths must be less than
$T'_W(r_{0\perp},p_{0\perp})$ given in Eq.~(\ref{eq:T'_W}).
Thus we see that
\begin{eqnarray}
& & \hskip -4mm \int_m \! \rmd {\bf Y}_1  \rmd {\bf Y}_3
\int_0^{T'_W} \rmd t_1 \rmd t_3
\langle P({\bf Y}_1,{\bf Y}_{01};t_1) P({\bf Y}_3,{\bf Y}_{03};t_3) \rangle 
\nonumber \\
& &
= {\rhom^2 \Nm p_{\rm F}^2 \cos \theta_{01} \cos \theta_{03} 
\over \sum_{m'} \rhomp (2-\rhomp)\Nmp} 
(1-\exp[-T'_W/\tDtwo]). \qquad \ 
\label{eq:prob-for-D_1a}
\end{eqnarray}
where $T'_W$ is the function of $(r_{0\perp},p_{0\perp})$ given in
Eq.~(\ref{eq:T'_W}).
Note that the denominator comes from the fact we are considering the survival
probability for the correlated paths; thus the probability that 
the correlation is destroyed
by paths escaping into a lead during the time $t$ to $t+ \de t$
is $P_2(t) \times \de t/\tDtwo$, where $P_2(t)$ is given by Eq.~(\ref{eq:P2}).
We insert Eq.~(\ref{eq:prob-for-D_1a}) into Eq.~(\ref{contribution}),
then just as in Section~\ref{sect:cbs0} we
change integration variables
using 
$p_{\rm F} \cos \theta _{03}\rmd {\bf Y}_{03} 
=\rmd r_{0\perp} \rmd p_{0\perp}$, and define 
$\tilde{p}_0\equiv p_{0\perp}+ m\lambda r_{0\perp}$.
In the regime of interest 
$T'_W(r_{0\perp}, p_{0\perp}) 
\simeq \lambda^{-1} \ln[m\lambda W/|\tilde{p}_0|]$.
Evaluating the integral over $r_{0\perp}$ leaves 
a $\tilde{p}_0$-integral which we cast as an Euler $\Gamma$-function, 
just as 
for coherent-backscattering in Section~\ref{sect:cbs0}.
To lowest order in $(\lambda\tDtwo)^{-1}$ we find the contribution to 
${\rm Tr}[{\mathbb S}_{mm_0}^\dagger {\mathbb S}_{mm_0}
{\mathbb S}_{mm_0}^\dagger {\mathbb S}_{mm_0}]$ shown in 
Fig.~\ref{fig:noise-trans1}a is
\begin{eqnarray}
D_{\rm 1a}
&=& {\rhomo^2 \rhom^2 \Nmo \Nm (1-\exp[-\tEo/\tDtwo])
\over \sum_{m'} \rhomp (2-\rhomp)\Nmp}   \, .
\label{eq:D1a}
\end{eqnarray}

Now we note that each contribution in Fig.~\ref{fig:noise-trans1} 
is of a similar form to $D_{\rm 1a}$.
For example, $D_{\rm 1b}$ and $D_{\rm 1c}$ 
are like  $D_{\rm 1a}$ with the exception that
a path is reflected off lead $m$ and then returns to lead $m$.
The result of the integral over
$(r_{0\perp}, p_{0\perp})$ is basically unchanged 
when we replace the action difference 
in Eq.~(\ref{eq:deltaS_cbs}) with Eq.~(\ref{eq:deltaS_cbs-indir}),
see Section~\ref{sect:cbs1}.
Thus we get each of 
these contributions by simply multiplying $D_{\rm 1a}$ by
\begin{eqnarray}
{1-\rhom \over \rhom }\times
{\rhom \Nm \over \sum_{m'}\rho_{m'}N_{m'}}  
&=& {(1-\rhom) \Nm \over \sum_{m'}\rho_{m'}N_{m'}},  
\label{eq:multiply-D_1a-by-this}
\end{eqnarray}
where after reflection the path that remains in the cavity
evolves alone, so its survival is governed by Eq.~(\ref{eq:P1}).
We can make the same argument
for paths which enter the cavity from lead $m_0$ at different times, 
but in such a way that the second enters the cavity at 
a moment when by chance the first is reflecting off barrier $m_0$, in such
a way that the paths form a pair.  To make the argument we need simply
reverse the direction of the paths, and we return to the situation discussed
above Eq.~(\ref{eq:multiply-D_1a-by-this}) with $m$ replaced by $m_0$.    
Thus we find that the sum of all contributions in Fig.~\ref{fig:noise-trans1}
is 
\begin{eqnarray}
D_1
&=& 
{\rhomo^2\rhom^2 \Nmo\Nm \big(1-\exp[-\tEo/\tDtwo]\big)
\over \sum_{m'} \rhomp (2-\rhomp) \Nmp}
\\
& & \times 
\left(1+ {2(1-\rhomo)\Nmo \over \sum_{m'} \rhomp \Nmp} \right)
\left(1+ {2(1-\rhom)\Nm \over \sum_{m'} \rhomp \Nmp} \right).
\nonumber 
\label{eq:D1}
\end{eqnarray}

\subsection{Noise contributions 
which vanish for infinite Ehrenfest time.}
\label{sect:D2}

To evaluate all the contributions in Fig.~\ref{fig:noise-trans2} except
Fig.~\ref{fig:noise-trans2}g, we note
that the path pairs are always correlated at one lead;
at that lead they only escape in an uncorrelated manner 
if one path tunnels while the other is reflected.
In this case the details of the
encounter are as given in Fig.~\ref{fig:cbs}, 
if we make the replacements given in Eq.~(\ref{eq:replacements}).
These contributions are extremely similar to those discussed in 
Section~\ref{sect:D1}, 
the only difference is that the paths become more than the
lead width apart at at least one lead.
This occurs because the paired paths survive for a time longer than $T'_W$. 
Thus for the contribution in Fig.~\ref{fig:noise-trans2}a,
we have
\begin{eqnarray}
& & \hskip -4mm \int_m \! \rmd {\bf Y}_1  \rmd {\bf Y}_3
\int_{T'_W}^\infty \rmd t_1 \rmd t_3
\langle P({\bf Y}_1,{\bf Y}_{01};t_1) P({\bf Y}_3,{\bf Y}_{03};t_3) \rangle 
\nonumber \\
& &
= {\rhom^2 \Nm^2 p_{\rm F}^2 \cos \theta_{01} \cos \theta_{03} 
\over (\sum_{m'} \rhomp \Nmp)^2} \; 
\exp[-T'_W/\tDtwo] \  
\, .\qquad \ 
\label{eq:prob-for-D_2a}
\end{eqnarray}
Since the paths separate inside the cavity under the classical dynamics,
their escape is given by the single path escape.
However the factor of $\exp[-T'_W/\tDtwo] $
is due to the fact that the correlated paths must first survive until time 
$T'_W(r_{0\perp}, p_{0\perp})$, and during this time the survival probability is given by
Eq.~(\ref{eq:P2}).
Compare this with the equivalent contribution, Eq.~(\ref{eq:prob-for-D_1a}), 
for pairs which escape on a time less than $T'_W$.
The integral over $(r_{0\perp}, p_{0\perp})$ can be evaluated in exactly the
same manner as $D_{\rm 1a}$ (and in the same manner as the coherent-backscattering
in Section~\ref{sect:cbs0}).
Then we find that
\begin{eqnarray}\label{D2a}
D_{\rm 2a}
&=& {\rhomo^2 \rhom^2 \Nmo \Nm^2 \exp[-\tEo/\tDtwo]
\over (\sum_{m'} \rhomp \Nmp)^2}   \, .
\label{eq:D2a}
\end{eqnarray}
We then note that contributions $D_{\rm 2b}$ and $D_{\rm 2c}$  
equal $D_{\rm 2a}$ multiplied by $(1-\rhomo)/(\sum_{m'} \rhomp\Nmp)$,
thus
\begin{eqnarray}
D_{\rm 2a} + D_{\rm 2b} + D_{\rm 2c}
&=& {\rhomo^2 \rhom^2 \Nmo \Nm^2 \exp[-\tEo/\tDtwo]
\over (\sum_{m'} \rhomp \Nmp)^2} 
\nonumber \\
& &\times
\left(1+ {2(1-\rhomo)\Nmo \over \sum_{m'} \rhomp \Nmp} \right)\, .
\label{eq:D2abc}
\end{eqnarray}

The contributions $D_{\rm 2d}$, $D_{\rm 2e}$ and $D_{\rm 2f}$ 
(shown in Fig.~\ref{fig:noise-trans2}d,e,f)
are the same as contributions $D_{\rm 2a}$, $D_{\rm 2b}$ and $D_{\rm 2c}$
(shown in Fig.~\ref{fig:noise-trans2}a,b,c)
with $m_0 \leftrightarrow m$.
Thus   
\begin{eqnarray}
D_{\rm 2d} + D_{\rm 2e} + D_{\rm 2f}
&=& {\rhomo^2 \rhom^2 \Nmo^2 \Nm \exp[-\tEo/\tDtwo]
\over (\sum_{m'} \rhomp \Nmp)^2} 
\nonumber \\
& &\times
\left(1+ {2(1-\rhom)\Nm \over \sum_{m'} \rhomp \Nmp} \right)\, .
\label{eq:D2def}
\end{eqnarray}

This leaves us to evaluate $D_{\rm 2g}$, the contribution shown in
Fig.~\ref{fig:noise-trans2}g,
this is different from all other noise contributions because the
paths are not correlated at escape.
Thus we evaluate this contribution in a manner similar to
the weak-localization contribution in Section~\ref{sect:wl0}.
We use the method developed by Richter and Sieber 
\cite{Ric02}, 
while taking into account
that paths in the same region of phase-space 
have escape probabilities given by Eq.~(\ref{eq:P2}).
This method was first applied to shot-noise in
Ref.~[\onlinecite{Haake-fano,wj2005-fano}] in the absence of tunnel-barriers. 
Here we follow Ref.~[\onlinecite{wj2005-fano}],
and we write the action difference as 
in Eq.~(\ref{eq:deltaS_wl0}),
where the crossing angle, $\eps$, is shown in Fig.~\ref{fig:noise-trans2}g.
We write
\begin{equation}
P({\bf Y}_i,{\bf Y}_{0i};t_i)= \!
\int \rmd {\bf R}_i 
\tilde{P}({\bf Y}_i,{\bf R}_i;t_i-t_i')P({\bf R}_i,{\bf Y}_{0i};t_i') \, ,
\nonumber
\end{equation} 
where $\tilde{P}$ is the probability for the classical path to exist 
(not multiplied by the injection momentum), and
${\bf R}_i$ is a point in the system's phase-space 
$({\bf r}_i,\phi_i)$ visited at time $t_i'$, 
with $\phi_i$ giving the direction of the momentum.
We choose ${\bf R}_1$ and ${\bf R}_3$ as the points at which the paths 
cross, so ${\bf R}_3 = ({\bf r}_1,\phi_1 \pm \eps)$ and
$\rmd {\bf R}_3 = v_{\rm F}^2 \sin \eps \rmd t_1' \rmd t_3' \rmd \eps$. 

To get $D_{\rm 2g}$ 
we sum only contributions where $\gamma1$ crosses $\gamma3$,
we then take twice the real part of this result to include the contributions
where $\gamma1$ and $\gamma3$ avoid crossing (and hence $\gamma2$ 
and $\gamma 4$ cross).
Thus 
\begin{eqnarray}
D_{\rm 2g} 
&=& 
2(2\pi \hbar)^{-2}  \int_{\rm L} \rmd {\bf Y}_{01} \rmd {\bf Y}_{03}
\nonumber \\
& \times & \int_0^\pi \rmd \eps \;
{\rm Re}\big[\e^{\rmi \de S_{D_1}/\hbar}\big] 
\big\langle I( {\bf Y}_{01}, {\bf Y}_{03};\eps) \big\rangle .
\label{eq:D_1-as-integral}
\end{eqnarray}
where $\de S_{D_1}$ is the same as $\de S_{\rm wl}$ 
in eq.~(\ref{eq:deltaS_wl0}).
The function $I( {\bf Y}_{01}, {\bf Y}_{03};\eps)$ 
is related to the probability that $\gamma3$ crosses $\gamma1$ 
at angle $\pm\eps$. Its average is independent of
${\bf Y}_{01,03}$, so 
$\langle I( {\bf Y}_{01}, {\bf Y}_{03};\eps) \rangle = 
\langle I(\eps)\rangle$.
Injections/escapes are more than $T_W(\eps)/2$ from the crossing, so
\begin{eqnarray}
\langle I(\eps)\rangle
\! &=& \! 
2v_{\rm F}^2 \; \sin \eps
\int_{\rm R} \rmd {\bf Y}_{1} \rmd {\bf Y}_{3} \int \rmd {\bf R}_1 \nonumber \\
& \times & \int_T^\infty \rmd t_1  
\int_{T/2}^{t_1-T/2} \! \rmd t_1' 
\int_T^\infty \rmd t_3 
\int_{T/2}^{t_3-T/2} \! \rmd t'_3
\nonumber \\
& \times & \,
\big\langle \tilde{P}({\bf Y}_1,{\bf R}_1;t_1-t_1') 
P({\bf R}_1,{\bf Y}_{01};t_1') 
\nonumber \\
& \times & \, 
\tilde{P}({\bf Y}_3,{\bf R}_3;t_3-t_3') 
P({\bf R}_3,{\bf Y}_{03};t_3') \big\rangle  \, ,
\quad \quad 
\end{eqnarray}
where $T_W$ is the function of $\eps$ given in Eq.~(\ref{eq:T_W}). 
We next note that
within $T_W/2$ of the crossing, paths $\gamma1$ and $\gamma3$
are so close
to each other that their joint survival probability is 
given by Eq.~(\ref{eq:P2}). 
Elsewhere $\gamma1,\gamma3$ escape
independently through either lead at any time, hence 
\begin{eqnarray}
\big\langle I(\eps) \big\rangle
&=&
{p_{\rm F}^4  \tDone \over \pi \hbar m }
{ \rhomo^2 \rhom^2 \Nm^2 
\cos \theta_{01}\cos\theta_{03} \sin \epsilon 
\over  (\sum_{m'} \rhomp \Nmp)^3} 
e^{-T_W(\eps)/\tDtwo} 
\nonumber \\
\end{eqnarray}
where we used $\Nmp=(\pi \hbar)^{-1}p_{\rm F}W_{m'}$, 
and assumed that the probability that $\gamma3$ is at 
${\bf R}_3$ at time $t_3'$ 
in a system of area $A$ is 
$(2\pi A)^{-1} =  m [2\pi\hbar \tau_{\rm D}\sum_{m'} \rhomp\Nmp]^{-1}$.
Then the ${\bf Y}_{01,03}$-integral in Eq.~(\ref{eq:D_1-as-integral})
gives $(2W_{m_0})^{2}$.  
The integral over $\eps$ is the same as given in eq.~(\ref{eq:eps-integral})
for weak-localization.
Thus
\begin{eqnarray}\label{eq:D2g}
D_{\rm 2g} &=& -{\rhomo^2 \rhom^2 \Nmo^2 \Nm^2 \sum_{m'}\rhomp(2-\rhomp) \Nmp
\over (\sum_{m'}\rhomp \Nmp)^4}
\nonumber \\
& & \times
 \exp[-\tEo/\tDtwo] \, .
\end{eqnarray}

Summing all the contributions in Fig~\ref{fig:noise-trans1},
given in Eqs.~(\ref{eq:D2abc},\ref{eq:D2def},\ref{eq:D2g}),
we get
\begin{eqnarray}
D_2
&=& 
{\rhomo^2\rhom^2 \Nmo\Nm \exp [-\tEo/\tDtwo]
\over (\sum_{m'} \rhomp \Nmp)^2} 
\nonumber \\
& & \times
\Bigg[ 
\Nmo+\Nm + {2(2-\rhomo-\rhom)\Nmo \Nm \over \sum_{m'} \rhomp \Nmp}
\nonumber \\
& & \qquad
- {\Nmo \Nm \sum_{m'} \rhomp (2-\rhomp) \Nmp
\over \big(\sum_{m'} \rhomp \Nmp\big)^2 }
\Bigg].
\label{eq:D2}
\end{eqnarray}

\subsection{Total shot-noise for arbitrary Ehrenfest time.}

Thus we find that the Fano factor is given by
\begin{eqnarray}
F&=& 1 - {(D_1 +D_2) \sum_{m'} \rhomp \Nmp \over \rhomo \rhom \Nmo \Nm}
\label{eq:Fano-result}
\end{eqnarray}
where $D_1$ and $D_2$ are given by Eqs.~(\ref{eq:D1},\ref{eq:D2}).

Let us first consider this result for transparent barriers 
($\rhomp = 1$ for all $m'$), in this case it simplifies to 
\begin{eqnarray}
F = \left[1 - {\Nmo + \Nm \over \sum_{m'} \Nmp} + 
{\Nmo\Nm \over (\sum_{m'} \Nmp)^2}\right] \exp [-\tEo/\tau_{\rm D}]
\nonumber \\
\label{eq:Fano-transparent}
\end{eqnarray}
where $\tau_{\rm D}^{-1}= (\tau_0 L)^{-1}\sum_{m=1}^n W_m$.
As noted in Ref.~[\onlinecite{wj2005-fano,Haake-fano}] 
for two leads without tunnel-barriers, the contributions to 
${\rm Tr}[{\mathbb S}_{mm_0}^\dagger {\mathbb S}_{mm_0}
{\mathbb S}_{mm_0}^\dagger {\mathbb S}_{mm_0}]$ 
which enter or exit the cavity in a correlated manner
cancel the the contribution of 
${\rm Tr}[{\mathbb S}_{mm_0}^\dagger {\mathbb S}_{mm_0}]$.
In that case the noise is given by the contribution which 
enters and exits the cavity in an uncorrelated manner 
(Fig.~\ref{fig:noise-trans1}g). 
Here we see this is only the case for two leads;
the first term in Eq.~(\ref{eq:Fano-transparent}) comes from
${\rm Tr}[{\mathbb S}_{mm_0}^\dagger {\mathbb S}_{mm_0}]$, while the
second term comes from those contributions to 
${\rm Tr}[{\mathbb S}_{mm_0}^\dagger {\mathbb S}_{mm_0}
{\mathbb S}_{mm_0}^\dagger {\mathbb S}_{mm_0}]$ 
which enter or exit the cavity in a correlated manner; 
in general these two do {\it not} cancel each other.
However we see that for an arbitrary number of leads without
 tunnel-barriers the paths shorter than the Ehrenfest time are noiseless
(just as for two leads\cite{wj2005-fano}).

Now let us consider eq.~(\ref{eq:Fano-result})
in the limit of opaque barriers
($\rhomp \to 0$ for all $m'$),
we see from Eq.~(\ref{eq:Fano-result}) that the
Fano factor is 
\begin{eqnarray}
F = 
1- {2 \rhomo \rhom 
\Nmo \Nm \over (\sum_{m'} \rhomp \Nmp)^2}, 
\label{eq:Fano-opaque1}
\end{eqnarray} 
which is completely {\it independent} of the Ehrenfest time.
In fact, when we expand Eq.~(\ref{eq:Fano-result}) to ${\cal O}[\rho]$,
we see that the ${\cal O}[\rho]$-term is also
independent of the Ehrenfest time, thus 
\begin{eqnarray}
F &=& 
1- {2 \rhomo \rhom (1 - \rhomo-\rhom)
\Nmo \Nm \over (\sum_{m'} \rhomp \Nmp)^2}
\nonumber \\
& &-  {\rhomo \rhom \Nmo \Nm \sum_{m'} \rhomp^2 \Nmp 
\over (\sum_{m'} \rhomp \Nmp)^3}
\nonumber \\
& & - {\rhomo \rhom (\Nmo+\Nm) \over \sum_{m'} \rhomp \Nmp}
+{\cal O}[\rho^2]
\label{eq:Fano-opaque2}
\end{eqnarray}
where the ${\cal O}[\rho^2]$-term is dependent on the Ehrenfest time.

In the case of a symmetric two lead cavity, 
$N_1=N_2=N$ and $\rho_1=\rho_2=\rho$,
the rather ugly result in Eq.~(\ref{eq:Fano-result}) 
greatly simplifies to
\begin{eqnarray}
F= {1-\rho\over 2-\rho}\big(1-\e^{-\tEo/\tDtwo}\big) + 
{2-\rho \over 4}  \e^{-\tEo/\tDtwo} \ .
\end{eqnarray} 
Both terms in this expression increase monotonically 
as we reduce the transparency of the tunnel-barriers from $\rho=1$ to $\rho=0$,
thus the presence of the tunnel-barriers always increases the quantum noise
(for given Ehrenfest time).
In the limit of transparent barriers ($\rho=1$),
the Fano-factor crosses over from the RMT value of $1/4$ to zero
as we increase the Ehrenfest time.
While in the limit of opaque barriers ($\rho\ll 1$)
the Fano factor is $1/2$, independent of the Ehrenfest time.

\subsection{Consistency check: alternative shot noise formula.}

Finally we note that
the main difficulty in the above calculation is the identification
of all contributions.  Thus it is important to have an independent check
that no contributions have been missed.  
In appendix~\ref{sect:noise-refl},
we use the fact that the Fano factor can be written in terms of
a product of transmission and reflection contributions, 
see Eq.~(\ref{eq:Fano-def-refl}),
and rederive the result in Eq.~(\ref{eq:Fano-result}) 
starting from that formula.
The transmission/reflection contributions calculated there 
(shown in Figs.~\ref{fig:noise-refl1} and \ref{fig:noise-refl2}) 
combine in a non-trivial manner to give the result 
in Eq.~(\ref{eq:Fano-result}).  
This can be seen by the fact that
there is no obvious $m_0 \leftrightarrow m$ symmetry
in the contributions in Appendix \ref{sect:noise-refl}
(compare that with the contributions calculated above, where
every contribution has a partner with   $m_0 \leftrightarrow m$).
Thus if we miss a contribution 
in Figs.~\ref{fig:noise-trans1},\ref{fig:noise-trans2}
and miss the equivalent contributions in 
Figs.~\ref{fig:noise-refl1},\ref{fig:noise-refl2},
it would be extremely unlikely that the two sets of contributions
would sum to give the same result.


\section{Concluding Remark.}

We have summarized the central physics discussed in this article
in Section~\ref{sect:qualitative}, and recommend that as a summary of 
the contents of this article.  Here we would like to re-iterate 
one technical point which makes the semiclassical calculations tractable.
It is the fact that when writing probabilities to escape into
a given lead we count only successful attempts to escape
(not situations where the particle hit the tunnel-barrier on a lead and 
was reflected back into the cavity).
This is very different from the equivalent RMT 
calculation\cite{Brouwer96-RMT-transport-diagrams} in which each collision
with the tunnel-barrier on a lead was explicitly taken into account,
regardless of whether the particle tunnels or not.
We feel unqualified to speculate if our approach could be directly
applied to simplifying the RMT calculations.  However we are certain
that the hand-waving arguments made in Section~\ref{sect:failed-cbs} 
would be an excellent
guide to the intuition when performing involved RMT calculations.

\section{Acknowledgments.}
I am grateful to M.~Polianski and C.~Petitjean for very useful 
and stimulating discussions, 
and thank Y.~Nazarov for drawing my attention to this problem.


\appendix

\section{Phase of transmission and reflection at a tunnel-barrier.}
\label{sect:barrier}

Here we calculate the complex transmission and reflection amplitudes,
$t$ and $r$, for
a rectangular tunnel-barrier, in the limit that the barrier is thin 
and high (width $l$ and height $U$).
We follow the standard procedure
explained in most quantum mechanics textbooks.
We include the details here simply because most textbooks
give the transmission and reflection probabilities, but not the phases.

We consider a one-dimensional system with the particle comes
from the left.  To the left of the tunnel-barrier
the solutions of Schr\"odinger's equation
have momentum $\pm p$, while to the right of the barrier 
the only non-zero solution has momentum $+p$.  
Inside the barrier the solutions decay/grow like
$\exp [\pm kx/\hbar]$, where $k=[2mU-p^2]^{1/2}$. 
Solving the simultaneous 
equations that we find by matching the wavefunction and 
its derivative at the two edges of the tunnel-barrier, we see that
the transmission amplitude is 
\begin{eqnarray}
t= 
{\exp [-\rmi pl/\hbar] \over 
\cosh (kl/\hbar) + {\rmi \over 2}(k/p-p/k) \sinh (kl/\hbar) },
\end{eqnarray}
while the reflection amplitude is
\begin{eqnarray}
r= -{\rmi\over 2}
{ (k/p + p/k) \sinh (kl/\hbar)
\over \cosh (kl/\hbar) + {\rmi \over 2} (k/p - p/k) \sinh (kl/\hbar) },
\end{eqnarray}
where $l$ is the width of the barrier.
To take the thin high barrier limit, 
we take $U\to \infty$ and $l\to 0$ such that $kl/\hbar$ remains finite.
We then adjust $kl/\hbar$ to give us the tunnelling probability,
$\rho=|t|^2$, that we wish;
in this limit $pl/\hbar=0$.

We will need  the transmission and reflection phases 
in Appendix~\ref{sect:noise-refl}, because for $\tilde{D}_{\rm 1b}$
we find that
$t$ and $r$ do not appear in the combination $|t|^2=\rho$ or $|r|^2=1-\rho$.
Instead we need to calculate the combinations $(t^*r)^2$ and $(r^*t)^2$.
From the above results for $t$ and $r$ we see that
\begin{eqnarray}
 [(e^{\rmi pl/\hbar}t)^* r]^2= [(e^{\rmi pl/\hbar}t) r^*]^2 
= - |t|^2|r|^2 = -\rho(1-\rho)
\qquad
\end{eqnarray}
and hence in the thin barrier limit ($pl \ll \hbar$),
we get $(t^*r)^2= (r^*t)^2 = - \rho(1-\rho)$.

We note that throughout this article, we assume 
that $\rho=|t|^2$ is unchanged as we take the classical limit 
$\lambda_{\rm F}/L \to 0$.  This means that we keep the ratio 
$\lambda_{\rm F}/l$ fixed 
(and do not scale $l$ like the classical scales $W,L$) 
as we take the limit $\lambda_{\rm F}/L \to 0$.
For simplicity we call this the {\it classical limit}. 
Now, one might argue that the strict classical limit 
would involve taking $\lambda_{\rm F}/l \to 0$, 
so there would be no tunnelling.  
However this limit is uninteresting because 
without tunnelling there would be no conduction at all.

\section{Approximating sum over lead modes by Dirac $\de$-function.}
\label{appendix:delta_hbar}

The approximation made above Eq.~(\ref{eq:conductance}) 
was introduced in Refs.~[\onlinecite{jw2005,wj2005-fano}], however the
explanation given there was too brief to be helpful.
The issue not discussed there was the fact that 
$\exp[\rmi S_\gamma/\hbar]$ in Eq.~(\ref{semicl-tr}) oscillates
as fast with $y$ as $\langle j|y\rangle$ does.
Thus we must deal explicitly with both fast oscillating terms at once.
To do this we write path $\gamma2$ (which goes to $y'$)
in terms of equivalent path that goes to $y$, then
$S_{\gamma2}(y') = S_{\gamma2} (y) +  \pF \sin \theta' (y'-y)
+ {\cal O}[m\lambda(y'-y)^2]$.
Hence when we write 
$\tr[{\mathbb S}_{mm_0}^\dagger{\mathbb S}_{mm_0}]$ 
we find that it contains 
$(y'-y)$-terms which oscillate fast 
(oscillate on a scale $\hbar/\pF=\lambda_{\rm F}$),
these terms are
$\sum_n\langle y'|n\rangle\langle n|y\rangle 
\exp[\rmi \pF \sin \theta' (y'-y)/\hbar]$.

Evaluating the sum for an ideal lead with $N$ lead modes of the form 
$\langle y|n\rangle= (2/W)^{1/2} \sin (\pi yn/W)$,
one finds that 
\begin{eqnarray}
& & \hskip -5mm
\sum_n\langle y'|n\rangle\langle n|y\rangle
\\
&=&{\sin [(z'-z)(N+1/2)] \over 2W\sin [(z'-z)/2]}
- {\sin [(z'+z)(N+1/2)] \over 2W\sin [(z'+z)/2]}
\nonumber 
\end{eqnarray}
where $z=\pi y/W$.
This function is peaked at $y'=y$
with width $\sim\lambda_{\rm F}$ and height $\sim\lambda_{\rm F}^{-1}$,
its integral over $y'$ is one.
Hence $\sum_n\langle y'|n\rangle\langle n|y\rangle \simeq \de(y'-y)$
for functions which varies slowly on the lengthscale
of a Fermi wavelength, where $\de(y')$ is a Dirac $\de$-function.
However the crucial point for this derivation is 
that one can also show that
$\sum_n\langle y'|n\rangle\langle n|y\rangle 
\exp[\rmi \pF \sin \theta' (y'-y)/\hbar] \simeq \de(y'-y)$.
To see this we consider the integral,
$ I = \int \rmd y' \sum_n\langle y'|n\rangle\langle n|y\rangle 
\exp[\rmi \pF \sin \theta' (y'-y)/\hbar]$. 
Defining $z = \pi(y'-y)/W $,
we find that  
  \begin{eqnarray}
  I &\simeq& (2\pi)^{-1} \int_{-\infty}^{\infty} {\rmd z \over z}
  \Big[ \e^{\rmi [N(1+\sin\theta')+1/2]z} 
  \nonumber \\
  & & \qquad \qquad - \e^{-\rmi [N(1-\sin\theta')+1/2]z}\Big], 
  \nonumber 
  \end{eqnarray}
where we use the fact that the integral is dominated by 
$z \sim N^{-1} \ll 1$. 
  The integrand is finite at $z=0$ even though the individual terms in the
  integrand diverge, so we can push the contour of integration 
  infinitesimally into the lower half of the complex plane.  
  To evaluate the integral over each term in the integrand, we note that
  the first (second) term converges in the upper (lower) half-plane.
  The first term's contour is deformed 
  into the upper-half plane, but then encircles
  the pole at $z=0$ giving the contribution $2\pi$. The second
  term's contour encircles nothing when pushed into the lower-half plane,
  so contributes nothing.  Thus $I=1$,
  independent of the prefactors in the exponents.
The integral is dominated by $z \sim N^{-1}$ (i.e.~$y \sim \lambda_{\rm F}$), hence 
$$
\sum_n\langle y'|n\rangle\langle n|y\rangle 
\exp[\rmi \pF \sin \theta' (y'-y)/\hbar] \simeq \de(y'-y), 
$$ 
for functions which vary slowly on the scale of the Fermi wavelength.


\begin{figure*}
\begin{center}
\hspace{-1cm}
\resizebox{14cm}{!}{\includegraphics{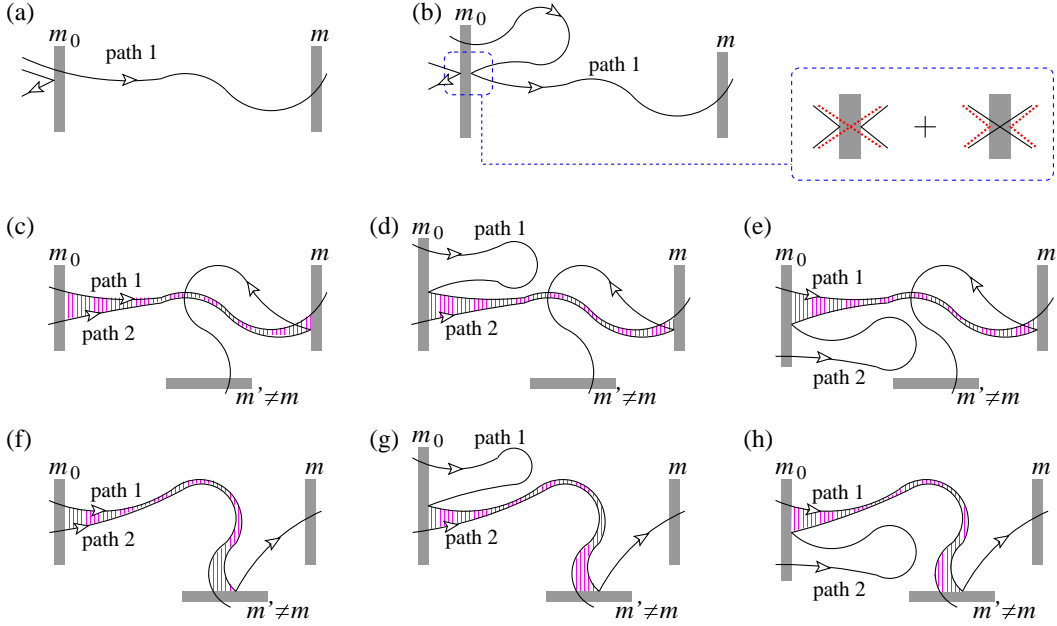}}
\caption{\label{fig:noise-refl1}
Here we list all the contributions to 
$\sum_{m'\neq m}{\rm tr}[{\mathbb S}_{m'm_0}^\dagger {\mathbb S}_{m'm_0}
{\mathbb S}_{mm_0}^\dagger {\mathbb S}_{mm_0}]$ 
which remain finite in the limit
of infinite Ehrenfest time.  The contributions are drawn in the same manner 
as Fig.~\ref{fig:noise-trans1}, with only paths 1 and 3 shown. 
The noise in these contribution is purely due to the stochastic nature of scattering at the tunnel-barriers, if they were absent these
contributions would be noiseless.  
The calculation will show that (a) and (b) are independent of $\tEo$, 
while all other contributions go like $(1-\e^{-\tEo/\tDtwo})$.
}
\end{center}
\end{figure*}

\section{Calculating Shot noise from transmission/non-transmission 
correlations.}
\label{sect:noise-refl}

The unitarity of the scattering matrix
means that 
\begin{eqnarray}
\sum_{m'=1}^n  [{\mathbb S}^\dagger]_{m'm_0} {\mathbb S}_{m'm_0}=1.
\end{eqnarray}
It follows that we can write the formula for the Fano-factor 
in Eq.~(\ref{eq:Fano-definition})
as
\begin{eqnarray}
F&=& 
{\sum_{m'\neq m}
{\rm Tr}[{\mathbb S}_{m'm_0}^\dagger {\mathbb S}_{m'm_0}
{\mathbb S}_{mm_0}^\dagger {\mathbb S}_{mm_0}] 
\over 
{\rm Tr}[{\mathbb S}_{mm_0}^\dagger {\mathbb S}_{mm_0}] }. \quad
\label{eq:Fano-def-refl}
\end{eqnarray}
Thus we see that the Fano factor is given by correlations
between transmitting contributions (those going to lead $m$)
and non-transmitting contributions (those going to any lead except $m$).

We now use the trajectory-method to calculate this directly.
We find no preservation of unitarity at the level
of the individual path's contributions to Eq.~(\ref{eq:Fano-definition})
and individual path's contributions to Eq.~(\ref{eq:Fano-def-refl}),
it is only when we have summed all contributions to 
Eq.~(\ref{eq:Fano-definition}) and Eq.~(\ref{eq:Fano-def-refl}),
that we find them to be equal.
We use this as a check on the fact we have included all relevant contributions,
and as a check on our algebra.

To calculate these contributions we use exactly the same method as
in Section~\ref{sect:noise}.  For a given contribution, one can take 
a contribution with the same topology in
Fig.~\ref{fig:noise-trans1} or Fig.~\ref{fig:noise-trans2}, and note
that one of the legs goes to $m'$ rather than $m$, with $m'$ being summed
over all leads {\it except} lead $m$.  
Thus here we simply give the results
for each contribution without discussing details of the calculation.
The exception to this is the contribution in Fig.~\ref{fig:noise-refl1}b
which has no analogous contribution in Section~\ref{sect:noise}.

The first two contributions in Fig.~\ref{fig:noise-refl1}  
are independent of the Ehrenfest time, they are
\begin{eqnarray}
\tilde{D}_{\rm 1a} 
&=&
{\rhomo (1-\rhomo) \rhom \Nmo\Nm \over 
\sum_{m'} \rhomp \Nmp}\, ,
\\
\tilde{D}_{\rm 1b} 
&=& -  
{2\rhomo^2 (1-\rhomo) \rhom \Nmo^2 \Nm \over 
(\sum_{m'} \rhomp \Nmp)^2}\, .
\end{eqnarray}
Note that the negative sign in $\tilde{D}_{\rm 1b}$ is due to the
tunnelling and reflection phases at the tunnel-barrier on
the $m_0$th lead.  Here, unlike for all other contributions, 
the complex tunnelling amplitude, $t$, and 
reflection amplitude, $r$, do {\it not} appear as products
of $|r|^2=\rho$ and $|t|^2= (1-\rho)$. 
Instead they appear in the combination $(t^*r)^2$ or $(tr^*)^2$,
thus we need the phases of $t,r$.
These are calculated in Appendix~\ref{sect:barrier},
where we show that for the high, thin barriers considered in this article
the phases generate an overall negative sign, so
$(t^*r)^2=(tr^*)^2= -\rho(1-\rho)$.

\begin{figure*}
\begin{center}
\hspace{-1cm}
\resizebox{14cm}{!}{\includegraphics{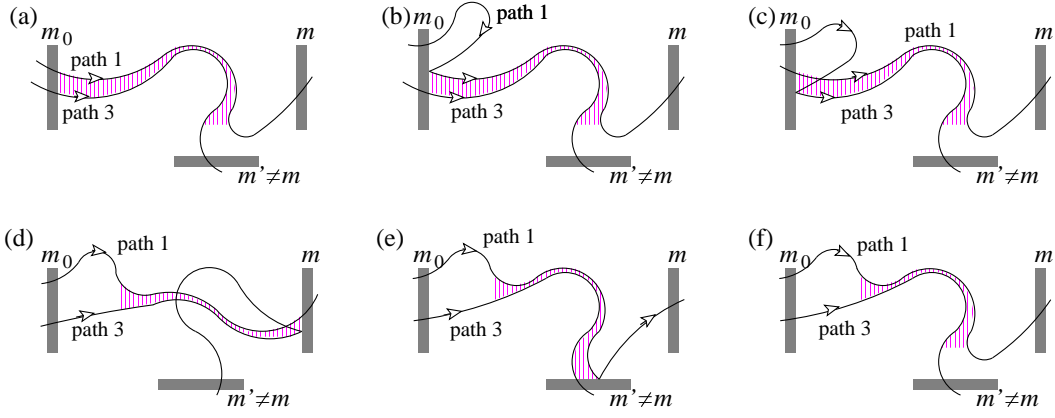}}
\caption{\label{fig:noise-refl2}
Here we list all the contributions to 
$\sum_{m'\neq m}{\rm tr}[{\mathbb S}_{m'm_0}^\dagger {\mathbb S}_{m'm_0}
{\mathbb S}_{mm_0}^\dagger {\mathbb S}_{mm_0}]$ 
which vanish in the limit of infinite Ehrenfest time.  
The contributions are drawn in the same manner as Fig.~\ref{fig:noise-trans1},
with only paths 1 and 3 shown. 
The calculation will show that all these contributions 
go like $\e^{-\tEo/\tDtwo}$.
Thus for $\tEo/\tDtwo \to 0$, the noise is given by the sum of 
these contributions and Fig.~\ref{fig:noise-refl2}a,b. 
}
\end{center}
\end{figure*}

All other terms in Fig.~\ref{fig:noise-refl1}  
go like $(1-\exp[-\tEo/\tDtwo])$.
The sum of the next three contributions in Fig.~\ref{fig:noise-refl1}  
gives
\begin{eqnarray}
\tilde{D}_{\rm 1(c+d+e)} 
&=&
{\rhomo^2 \rhom \Nmo\Nm (1-\exp[-\tEo/\tDtwo] )
\over 
(\sum_{m'} \rhomp (2-\rhomp) \Nmp)(\sum_{m'} \rhomp \Nmp) }
\nonumber \\
& &\times
\left( \sum_{m'} \rhomp (1-\rhomp)\Nmp - \rhom (1-\rhom)\Nm \right)
\nonumber \\
& &\times
\left( 1+ { 2(1-\rhomo)\Nmo \over \sum_{m'} \rhomp \Nmp } \right) \, .
\end{eqnarray}
The sum of the final three contributions in Fig.~\ref{fig:noise-refl1}  
gives
\begin{eqnarray}
\tilde{D}_{\rm 1(f+g+h)} 
&=&
{\rhomo^2 \rhom \Nmo\Nm (1-\exp[-\tEo/\tDtwo] )
\over 
(\sum_{m'} \rhomp (2-\rhomp) \Nmp)(\sum_{m'} \rhomp \Nmp) }
\nonumber \\
& & \hskip -2mm \times
\left( (1-\rhom)\sum_{m'} \rhomp \Nmp - \rhom (1-\rhom)\Nm \right)
\nonumber \\
& &\times
\left( 1+ { 2(1-\rhomo)\Nmo \over \sum_{m'} \rhomp \Nmp } \right) \, .
\end{eqnarray}
The sum of all the contributions shown in Fig.~\ref{fig:noise-refl1},
gives the noise in the infinite Ehrenfest time limit (when all
contributions in Fig.~\ref{fig:noise-refl2} are zero), and is
in agreement with the equivalent result, $D_1$, in Section~\ref{sect:noise}.

We now turn to the contributions in Fig.~\ref{fig:noise-refl2}.
The sum of the first three contributions in Fig.~\ref{fig:noise-refl2}
gives
\begin{eqnarray}
& & \hskip -5mm 
\tilde{D}_{\rm 2(a+b+c)} =
\\& &
{\rhomo^2 \rhom  \Nmo\Nm \exp[-\tEo/\tDtwo ] 
\over \sum_{m'} \rhomp \Nmp} 
\nonumber \\
& & \times
\left( 1- {\rhom \Nm \over \sum_{m'} \rhomp \Nmp } \right) 
\left( 1+ { 2(1-\rhomo)\Nmo \over \sum_{m'} \rhomp \Nmp } \right) \, .
\nonumber
\end{eqnarray}
The other three contributions in Fig.~\ref{fig:noise-refl2}
are
\begin{eqnarray}
\tilde{D}_{\rm 2d} 
&=&
{\rhomo^2 \rhom  \Nmo^2\Nm \exp[-\tEo/\tDtwo ] 
\over (\sum_{m'} \rhomp \Nmp)^3} 
\\
& & \times
\left(\sum_{m'} \rhomp (1-\rhomp)\Nmp -\rhom (1-\rhom)\Nm \right) \, ,
\nonumber 
\\
\tilde{D}_{\rm 2e} 
&=&
{\rhomo^2 \rhom  \Nmo^2\Nm \exp[-\tEo/\tDtwo ] 
\over (\sum_{m'} \rhomp \Nmp)^3} 
\\
& & \times 
\left( (1-\rhom)\sum_{m'} \rhomp \Nmp - \rhom (1-\rhom)\Nm \right) \, ,
\nonumber 
\\
\tilde{D}_{\rm 2f} 
&=& - {\rhomo^2 \rhom  \Nmo^2\Nm \exp[-\tEo/\tDtwo ] 
\over (\sum_{m'} \rhomp \Nmp)^3}  
\\
& &  
\times \left(1- {\rhom \Nm \over \sum_{m'} \rhomp \Nmp} \right)  
\sum_{m'} \rhomp(2-\rhomp) \Nmp \, .
\nonumber 
\end{eqnarray}
The sum of all the contributions shown in Fig.~\ref{fig:noise-refl2}
{\it and} the contributions shown in Fig.~\ref{fig:noise-refl1}a,b
gives the noise in the zero Ehrenfest time limit (when all
other contributions in Fig.~\ref{fig:noise-refl1} are zero), and is
in agreement with the equivalent result, $D_2$, in Section~\ref{sect:noise}.

Substituting the sum of all contributions in Figs.~\ref{fig:noise-refl1} and 
\ref{fig:noise-refl2} into the numerator of Eq.~(\ref{eq:Fano-def-refl}),
we again arrive at the result given in Eq.~(\ref{eq:Fano-result}). 
This shows that our method respects the unitarity 
of the scattering matrix (thereby conserving current), 
and acts as a check on the algebra.

\bibliographystyle{apsrev}

\begin{thebibliography}{99}
\bibitem{caveat2}  
We neglect the lead asymmetry 
in the definition of $T_W(\epsilon)$.
We assume $\ln[W'/W''] \sim 1$, so 
the assumption is good to zero order in $(\lambda \tDtwo)^{-1}$.
\end{thebibliography}

\end{document}